\documentclass[a4paper,preprintnumbers,showpacs,twocolumn,superscriptaddress,nofootinbib,amsmath,amssymb]{revtex4-1}
\usepackage{amsmath,amssymb,graphicx,amsfonts}
\usepackage{mathrsfs}
\usepackage{comment}
\usepackage{xcolor}
\usepackage[shortlabels]{enumitem}
\usepackage[unicode, colorlinks=true, linkcolor=magenta, citecolor=cyan, filecolor=cyan, urlcolor=blue, linktocpage, breaklinks]{hyperref}
\usepackage{url}
\usepackage{graphicx}
\usepackage{dcolumn}
\usepackage{bm}
\usepackage{wasysym}
\usepackage{color}
\usepackage{multirow}
\usepackage{booktabs}
\usepackage{float}

\begin{document}
\title{Effect of the charge asymmetry and orbital angular momentum in the entrance channel on the hindrance to complete fusion}

\author{Elzod Khusanov}
\email{xusanovelzod.99@gmail.com}
\affiliation{Institute of Nuclear Physics, Tashkent 100214, Uzbekistan}
\affiliation{Faculty of Physics, National University of Uzbekistan, Tashkent, Uzbekistan}
\affiliation{Institute of Theoretical Physics, National University of Uzbekistan, Tashkent 100174, Uzbekistan}
\author{Avazbek Nasirov}
\email{nasirov@jinr.ru} 
\affiliation{Institute of Nuclear Physics, Tashkent 100214, Uzbekistan}
\affiliation{Joint Institute for Nuclear Research, Dubna 141980, Russia}
\author{Mukhtorali Nishonov}
\affiliation{Faculty of Physics, National University of
Uzbekistan, Tashkent, Uzbekistan}

\begin{abstract}
 The hindrance to complete fusion is studied as a function of the charge asymmetry of colliding nuclei and orbital angular momentum of the collision. The formation and evolution of a dinuclear system (DNS)  in the  heavy ion collisions at energies  near the Coulomb barrier is calculated in the framework of the DNS model. The DNS evolution is considered as nucleon transfer between its fragments. The results prove that a hindrance at formation of a compound nucleus (CN) is related with the quasifission process which is breakup of the DNS into products instead to reach the 
 equilibrated state of the CN. The role of the angular momentum in the charge (mass) distribution of the reaction products for the given mass asymmetry of the colliding nuclei has been demonstrated. The results of this work have been compared with the measured data for the quasifission yields in the $^{12}$C+$^{204}$Pb and $^{48}$Ca+$^{168}$Er reactions to show the role of the mass asymmetry of the entrance channel. 
\end{abstract}

\maketitle
\section*{Introduction}

One of the problems of modern physics is for the synthesis of the extremely heavy chemical elements, therefore, the investigation of the target and projectile pair and corresponding range of the beam energy leading to as possible large cross sections of the evaporation residues is an important aim of researchers of nuclear physics.

The experimental and theoretical studies of the peculiarities of the processes occurring in heavy ion collisions are important to establish complete fusion mechanism. It can be done by the analysis of the observed reaction products. The absence of the full understanding the reaction mechanisms is related with difficulties of the unambiguous identifications of the mechanisms which are responsible for the  yield of the corresponding observed reaction products. There is a probability of the overlap of the mass distributions of the contributions from the two mechanisms: for example, the quasifission and  fusion-fission mass distributions may overlap in the mass asymmetric part of the yields \cite{Thakur2017,Atreya2023}. 

\begin{figure*}[ht]
	\hspace{1.5 cm}
	\includegraphics[width=0.80\textwidth]{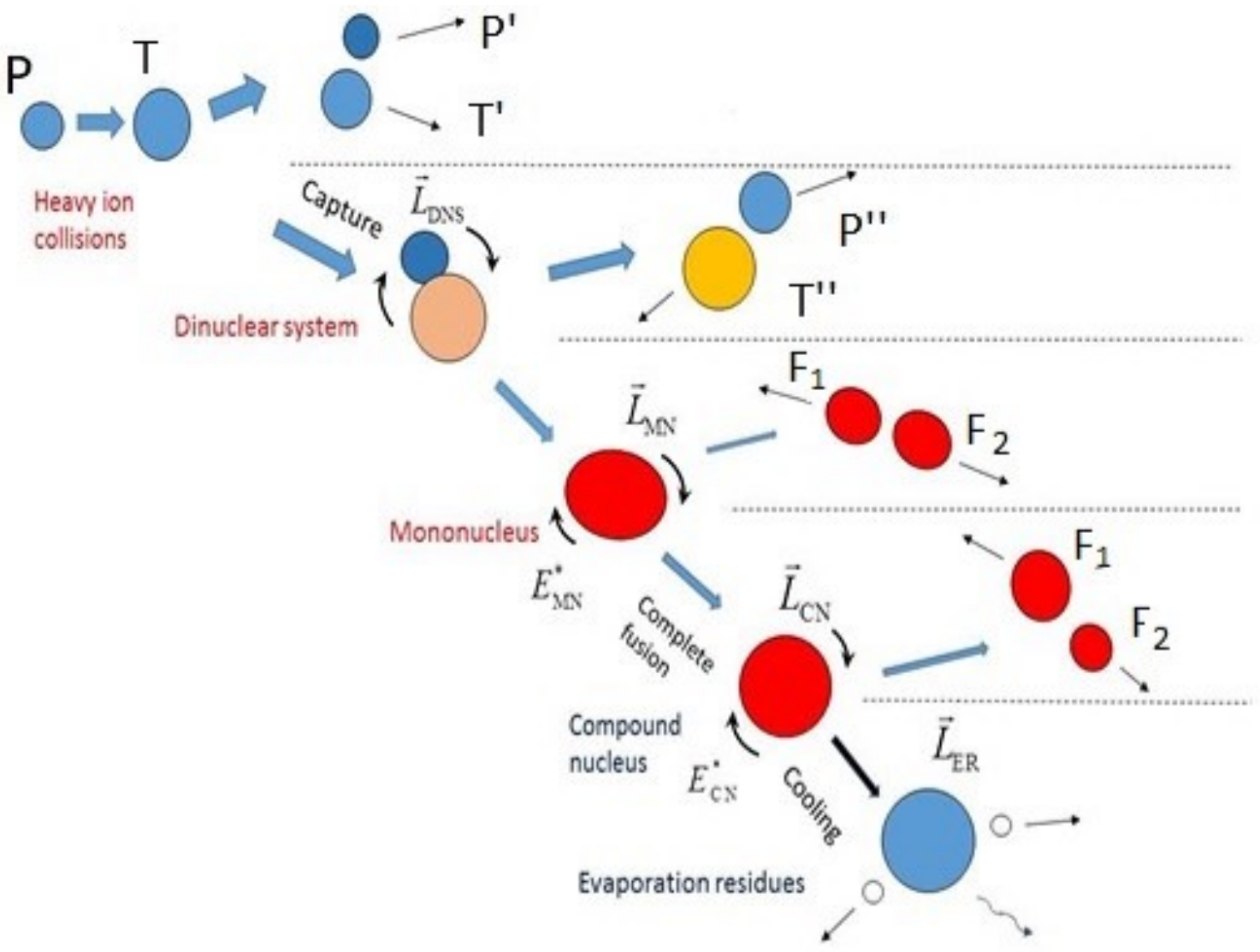}
\caption{(Color on-line) The sketch of the sequences of nuclear reaction channels which overcome competitions in different stages of interaction processes of dinuclear system fragments.} 
\label{channels}
\end{figure*}

Therefore, the analysis of the yields of the quasifission products allows us to  study the nature of a hindrance in the complete fusion. In the experiments on the setup CORSET of the Flerov Laboratory of nuclear reactions (JINR) \cite{Chizhov2003}, the fission-like binary products of the processes (fusion-fission, quasifission and fast fission) are registered by the two-arm time-of-flight spectrometer CORSET  by the  coincidence method of simultaneous recording. Naturally, the products of these binary processes can arrive to the same  detector with different probabilities. The mass and energy distributions of fission fragments were studied on the setup CORSET for the two reactions $^{12}$C+$^{204}$Pb and $^{48}$Ca+$^{168}$Er that lead to the same CN $^{216}$Ra$^*$ \cite{Chizhov2003}. The beam energies were fixed that the excitation energy of the being formed CN was around 40 MeV in both cases. The analysis of the measured mass and energy distributions showed that the contribution from asymmetric fission in the first reaction is only around 1.5\% but is about 30\% in the second. The authors have interpreted this dramatic increase in the asymmetric yield as a manifestation of the quasifission  process related with the shell effects for the reaction with $^{48}$Ca. They stressed that the more mass symmetric colliding nuclei in the entrance channel and high angular momentum populated in the reaction with $^{48}$Ca will clearly facilitate the evolution of the DNS toward the favored quasifission mass partition. The mass and charge distribution of the quasifission products may overlap with the ones of the fusion-fission and the deep-inelastic collisions \cite{DiazTorres2001,Adamian2003}.  The last process produces binary products with the mass and charge numbers around the values of the ones of the projectile- and target-nuclei. The overlap of the mass and/or angular distributions of the quasifission and fusion–fission products causes ambiguity in the estimation of the experimental fusion cross sections. But it is difficult  to separate them by the experimental methods as products of the corresponding processes. It is important to establish theoretically contributions in the yield of the reaction products from the  different mechanisms. 

 There are different theoretical models to describe the experimental data of the fusion cross sections, but there is not an unambiguous conclusions about fusion mechanism. The models based on the DNS concept consider complete fusion as multinucleon transfer from the 
 light nucleus of the DNS to its heavy one as diffusion process 
 \cite{Adamian2003,Nasirov2005,WangNan2012,BaoXJ2018,Rana2021,Cap2022}.
 The  ER formation is directly related with the fusion mechanism and ER products are registered enough unambiguously. Therefore, theoretical results are aimed to be close to the experimental data of evaporation residues. But the contribution of the ER yields is not alone providing the cross section of the CN formation in the complete fusion. The CN can undergo to fission into two (or three) fragments. The probability of fission increases by increasing its charge number $Z_{\rm CN}$, excitation  energy $E^*_{\rm CN}$ and angular momentum ($L_{\rm CN}$). The fusion mechanism is studied by the analysis of the dependence of the cross section of the complete fusion on the parameters of the reaction entrance channel as the charge (mass) asymmetry of colliding nuclei, the orientation angles of their axial symmetry axis, colliding energy and orbital angular momentum 
 \cite{Nasirov2005}. Consequently,  the fusion cross section is determined as a sum $\sigma_{\rm fus}=\sigma_{\rm ER}+\sigma_{\rm fiss}$. In the reaction leading to formation of the actinides $(Z>92)$ the fission process is dominant against ER formation. The experimental study of the complete fusion may be not unambiguous due to the presence of the
 contributions of binary fragments formed in the other channels of reaction in the cross 
 section $\sigma_{\rm fiss}$ of the fusion-fission products.  One of them is quasifission process which the breakup of DNS before reaching the CN. 
Fig. \ref{channels} shows the reaction channels producing binary products, which 
are observed in the experiments. It should be noted the difference between the deep-inelastic collision and quasifission process. The quasifission process is related with the capture events where full momentum transfer of the relative motion of colliding nuclei takes place. The  fusion and quasifission processes are two alternative processes of the capture reactions: the increase of the quasifission yields causes the decrease of the complete fusion events $\sigma_{\rm cap}=\sigma_{\rm fus}+\sigma_{\rm qf}$. Therefore, the investigation of the quasifission yields allows us to study the change of the intensity of the complete fusion events as a function of the entrance channel parameters.The hindrance to complete fusion is studied as a function of the charge asymmetry of colliding nuclei and orbital angular momentum of the collision \cite{Kayumov2022,Nasirov2024}. 

  The branching ratios between the realization probability of the different channels  depend on the mass and charge numbers  of the projectile and target nuclei  and kinematic parameters of the collision \cite{Kayumov2022}. In collisions with the large values $L> L_{gr}$ orbital momentum $L$ elastic and  inelastic scattering take place. The capture of colliding nuclei is a necessity condition for the CN formation. But this stage competes with the quasifission which produces binary products ($P''$ and $T''$). The quasifission products may have characteristics similar to the ones of the fission products. The CN stability is determined by its excitation energy $E^*_{\rm CN}$ and angular momentum $L_{\rm CN}$ since the fission barrier $B_f$ is a function of $E^*_{\rm CN}$ and $L_{\rm CN}$. If the being formed CN has angular momentum $L$ which is larger than the value $L_f$ causing completely disappearance of the fission barrier $B_f$ the system undergoes 
to the fast fission producing fragments ($F_1$ and $F_2$). It occurs only in collisions with $L \ge L_f$.
 
The DNS survived against to quasifission and fast fission  is transformed to the rotating and heated CN. If it survives  against fission  during cooling (de-excitation cascade), evaporation residue nucleus is formed. The contribution of the quasifission against complete fusion and the contribution of the fusion-fission of CN against its surviving by neutron emission are increased at the CN formation with large charge numbers $Z>92$. Therefore, the cross-section of syntheses of superheavy elements (SHE) can reach very small values. 

 In the case of the collision of the light nuclei with  the target nucleus capture 
 can be considered as the complete fusion since the intrinsic barrier $B^*_{\rm fus}$ 
 causing a hindrance to complete fusion is small. 
 But theoretical investigation of the  yield of the binary reaction products observed in the mass symmetric and  mass asymmetric entrance channels of the reactions, as well as the study  of the hindrance to the  complete fusion leading to the formation of the  superheavy elements show  that there is a large difference between capture and complete fusion cross sections in case of collision of the massive nuclei. The hindrance to complete fusion in reactions with massive nuclei is explained by the presence of an internal barrier $B^*_{\rm fus}$ associated with internal structural effects in DNS fragments \cite{Nasirov2005,Nasirov2024}. The value of  $B^*_{\rm fus}$ depends on the characteristics of projectile and target nuclei in the entrance channel and orbital angular momentum. During the development of the resulting DNS, its fragments may separate relatively early before reaching the CN state.

In Section 2, the basic physical quantities as potential energy surface (PES), intrinsic fusion barrier, quasifission barrier,  evolution of the DNS charge (mass) asymmetry are described.  Discussion of the results of this work and comparison with the corresponding experimental data are presented in Section 3.

\section{Theoretical formalism}

In this work, the range of the values of the orbital angular momentum leading to capture is 
determined by solving the dynamical equations of motion for the relative distance $R$ and 
orbital angular momentum $L$ \cite{Fazio2003,Nasirov2005,Nasirov2019}. The contributions 
coming from the breakup (quasifission) of the DNS formed in the different angular momentum 
$L=\ell \hbar$ are included into consideration by the expression:
\begin{eqnarray}
    Y_Z(E_{\rm c.m.}, \alpha_i)&=&\sum\limits_{\ell=0}^{\ell_d}(2\ell+1)\mathcal{P}_{\rm cap}(E_{\rm c.m.},\ell, \alpha_i)\nonumber\\
     &\times&Y_Z(E_{\rm c.m.},\ell, \alpha_i),
\end{eqnarray}
where $\mathcal{P}_{cap}(E_{c.m.},\ell, \alpha_i)$ is the capture probability 
for the colliding nuclei with the orientation angles $\alpha_i$ ($i=1,2$) of the 
axial symmetry axis relative to the beam direction (for the deformed nuclei,
see Fig. \ref{Burchak} in Appendix); 
 $Y_Z (E_{\rm c.m.},\ell)$ is  the  probability of the yield of the fragment with the charge number $Z$ in the collision with the energy $E_{\rm c.m.}$ and orbital angular momentum $\ell$; $\ell_d$ is the maximum value of the orbital angular momentum leading to the capture (full momentum transfer of the relative motion) process. It is calculated by the solution of the dynamical equations of the relative motion and angular momentum $\ell$ 
\cite{Fazio2003,Nasirov2005}.   If the shape of nuclei in their ground state is spherical, during interaction they are deformed due to the surface vibration \cite{Nasirov2023PLB}. 
For excited states, quadrupole $2^+$ and octupole $3^-$ deformation parameters of the nuclei are assumed to be equal to their vibrations. deformation parameters (quadrupole 
$\beta^{(i)}_2$ and octupole $\beta^{(i)}_3$)  used for the DNS fragments ($i=1,2$).
Deformation parameters for excited states are obtained from $\beta_2^+$ \cite{Raman2001} and $\beta_3^-$ \cite{Spear1989}.

 The capture probability depend on the beam energy, the size of the potential well of the nucleus-nucleus potential, and the friction coefficients of radial motion and angular momentum. The size of the potential well and friction coefficient determine the number of the partial waves ($L=\ell \hbar$) leading to the capture of the projectile-nucleus by the target-nucleus. The size of the potential well is sensitive to the charge and mass asymmetry of the colliding nuclei. This fact has been demonstrated in Ref. \cite{Nasirov2019} by comparison of the nucleus-nucleus potential calculated for the $^{36}$S+$^{206}$Pb and $^{34}$S + $^{208}$Pb reactions. The nucleus-nucleus interaction potential, radial and tangential friction coefficients and inertia coefficients are calculated in the framework of the DNS model \cite{Fazio2003,Nasirov2005,Nasirov2019}.

\begin{figure}[ht]	
	\begin{center}
		\includegraphics[scale=0.3]{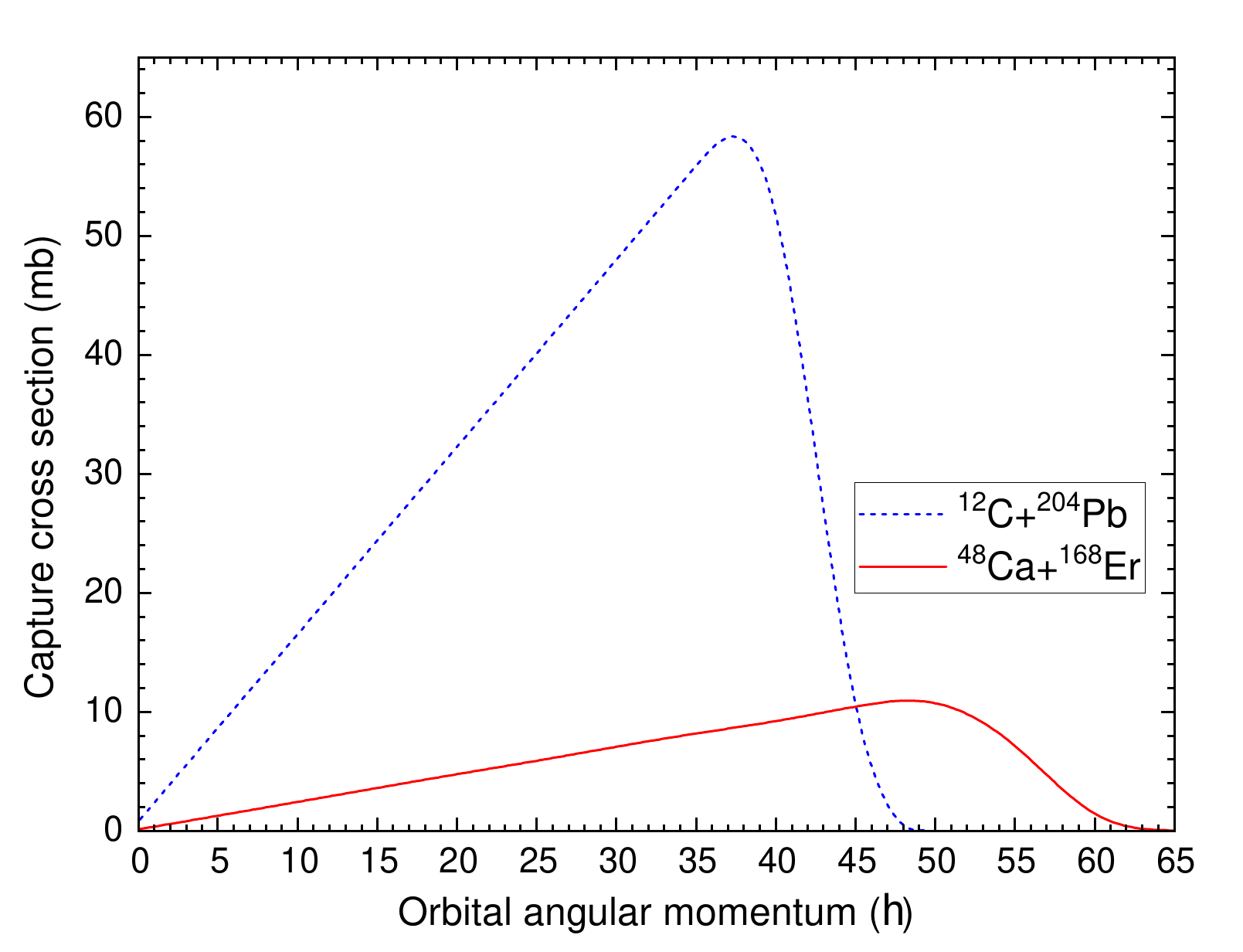}
		\caption{Comparison of the capture cross sections calculated in this work for the 
  $^{12}$C+$^{204}$Pb and  $^{48}$Ca+$^{168}$Er reactions.} 
		\label{CompCap} 
	\end{center}
\end{figure}

In Fig. \ref{CompCap}, the partial cross sections $\sigma_{\rm cap}^{(\ell)}$ calculated for the capture process in the $^{12}$C+$^{204}$Pb and  $^{48}$Ca+$^{168}$Er reactions for the energies $E_{\rm Lab}$=73 MeV and 153 MeV,   respectively, are compared. These energies correspond to the CN excitation energies $E_{\rm Lab}$=40.4 MeV  and 39.6 MeV for the corresponded reactions.  The critical values $L_{\rm cr}$ of the angular momentum estimated 
by the authors of Ref. \cite{Chizhov2003} for the $^{12}$C+$^{204}$Pb and  $^{48}$Ca+$^{168}$Er reactions were equal to $31 \hbar$ and $54 \hbar$, respectively. 
These values of $L_{\rm cr}$ are close to the  orbital angular momentum corresponding to the 
maximal values of the partial capture cross sections presented in Fig. \ref{CompCap}. 
The  values of $L_{\rm cr}$ obtained in  Ref. \cite{Chizhov2003} are obtained for the triangle
shape of the partial capture cross section which has sharp decrease at $L=L_{\rm cr}$.
The slow decrease of the theoretical curves of the partial capture cross section at large 
values of $L$ 
 in this work is related by the averaging of the results obtained 
for the collisions with different orientation angles $\alpha_i$ ($i=1,2$).

 \begin{figure*}[ht]
	\hspace{-2.5 cm}
	\includegraphics[width=1.0\textwidth]{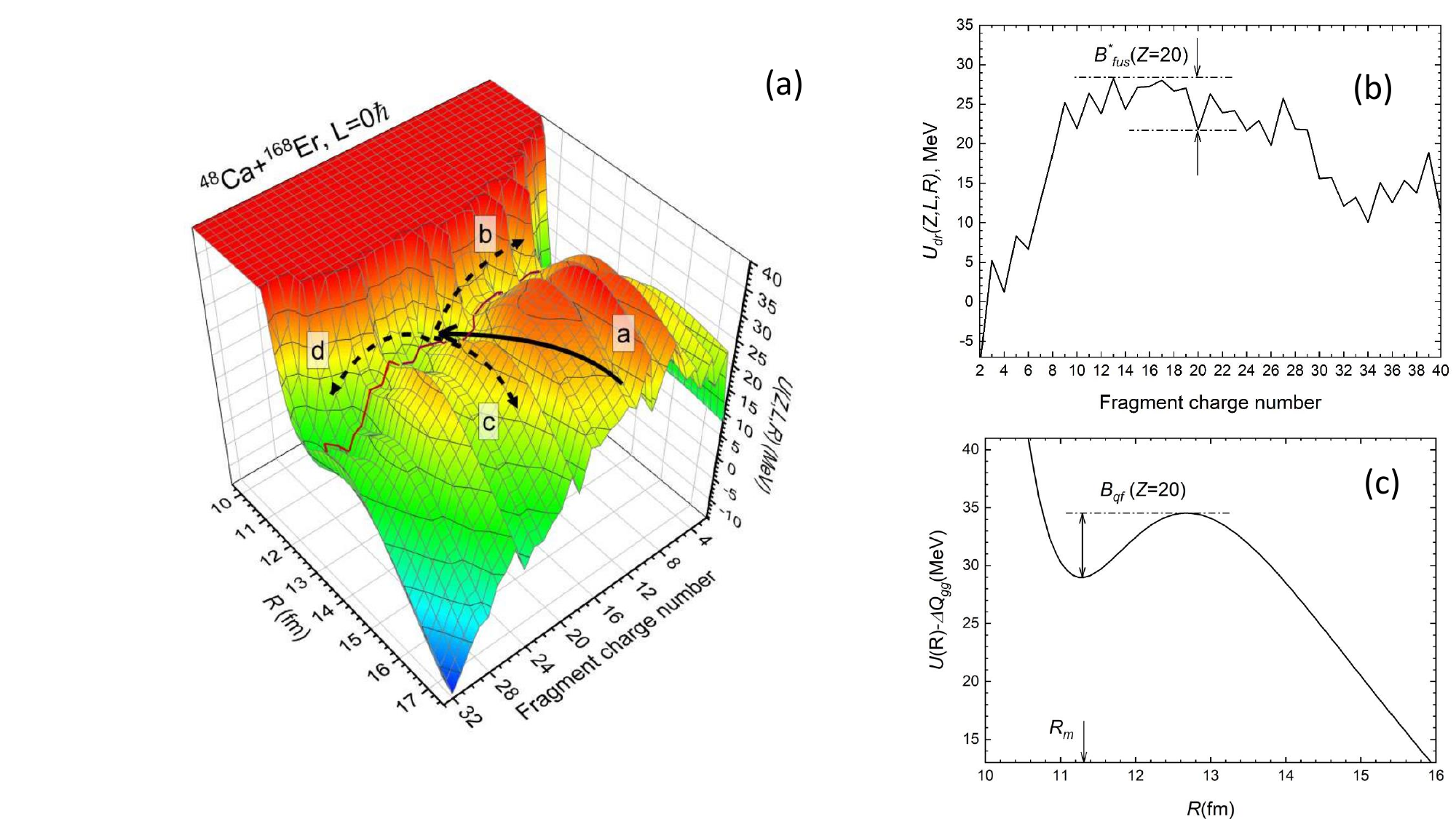}
\caption{(Color online).  Potential energy surface calculated for the DNS formed in the $^{48}$Ca+$^{168}$Er reaction as a	function of the charge number  ($Z$) of its fragment and relative distance ($R$) between centres-of-mass fragments: arrow (a) shows capture trajectory as the decrease 
of the kinetic of the relative motion; arrow (b) is direction of the complete fusion
due to nucleon transfer from the light fragment of the DNS to the heavy one;  arrows (c) 
and  (d) are the quasifission trajectories leading formation of the products with the 
different mass numbers (a). The driving potential of the DNS formed in the $^{48}$Ca+$^{168}$Er reaction as a function of the charge number ($Z$) of its light fragment; 
the intrinsic fusion $B^*_{\rm fus}$ barrier is determined for the entrance channel $Z=20$ (b). Quasifission $B_{\rm qf}$ barrier  of the entrance channel $Z=20$
 calculated as the depth of the potential well of the nucleus-nucleus interaction(c). } 
\label{PES2barr}
\end{figure*}

It is seen that partial cross section the $^{12}$C+$^{204}$Pb reaction is sufficiently larger than that for the $^{48}$Ca+$^{168}$Er reaction. The large values of $\sigma_{\rm cap}^{(\ell)}$ for the former reaction is explained with the fact it has small reduced mass $\mu=A_P A_T/(A_P+A_T)$=11.3 MeV  while it is equal to 37.3 MeV for the last reaction. Here $A_P$ and  $A_T$ are mass numbers of the projectile- and target-nuclei, respectively. Further evolution of the DNS is determined by the landscape of the potential energy surface calculated for the considered reactions.

\subsection{ Potential energy surface}

In the DNS approach, the PES plays a crucial role in theoretical study of the competition between complete fusion and quasifission processes which occur due to the multinucleon transfer between fragments of the DNS formed at capture. The PES represents  the total energy of the DNS as a function of its charge asymmetry $Z$ and relative distance $R$ between the mass-of-centres its interacting fragments. The landscape of the PES determines the fusion probability and yields of the quasifission products as a function of the beam energy and initial orbital angular momentum. The PES is calculated as a sum of the energy balance $Q_{gg}$ and nuclear-nuclear interaction potential $V(Z,A,L,R)$: 
\begin{eqnarray}
U(Z,A,L,R,\{\alpha_i\})=Q_{gg}+V(Z,A,L,R,\{\alpha_i\})\label{PES}
\end{eqnarray}
$Q_{gg}=B_1+B_2-B_{\rm CN}$ is the energy balance of the reaction, $B_1,~B_2$ and $B_{CN}$ are the interacting nuclei and the binding energies of CN taken from the table in Refs. \cite{AUDI2003,Moller1995}; the interaction potential $V$ is a sum of the Coulomb $V_{\rm Coul}$, nuclear $V_{\rm nuc}$   and rotational $V_{\rm rot}$ parts:
\begin{eqnarray}
V(Z,A, L,R,\{\alpha_i\})&=&V_{\rm Coul}(Z,A,L,R,\{\alpha_i\})\cr&+&
V_{\rm rot}(Z,A,L,R,\{\alpha_i\})\cr &+&V_{\rm nuc}(Z,A,L,R,\{\alpha_i\}).
 \label{Vint}
\end{eqnarray} 
The expression of the given potentials are presented in Appendix A. Here $Z_c=Z_{\rm CN}-Z$ and $A_c=A_{\rm CN}-A$ are introduced to mark the charge and mass numbers for the conjugate nucleus, respectively.  $A_{\rm CN}=A_P+A_T$  and $Z_{\rm CN}=Z_P+Z_T$, where  $Z_P$ and $Z_T$
are the charge numbers of the projectile and target nuclei, respectively.

 At large distances, the electrostatic repulsion between the positively charged nuclei dominates in the PES. The potential barrier appears at the distances $R\simeq R_P+R_T+2$ fm due to the competition between the nuclear and Coulomb forces. The driving potential is determined from the PES by taking  the values of the relative distance $R_m$ corresponding to the minimum of the potential well of the nucleus-nucleus interaction for the wide range of the charge numbers of the DNS fragments \cite{Nasirov2005}:
\begin{eqnarray}
U_{dr}(Z,A,L,R_m,\{\alpha_i\})=Q_{gg}+V(Z,A,L,R_m,\{\alpha_i\})\label{drivpot}
\end{eqnarray}

The competition between complete fusion and quasifission for the given charge and mass number of the DNS light fragment  is determined by the heights of the intrinsic fusion $B_{\rm fus}^*$  and quasifission $B_{qf}$ barriers \cite{Nasirov2005}. Their values depend on the angular momentum since PES is a function of $L$. As the nuclei approach each other, the PES changes shape, becoming more complex and exhibiting multiple minima and maxima as a function of its charge asymmetry which changes the binding energies $B_1$ and $B_2$ 
 of the DNS  fragments.
\begin{figure}[ht]
		\begin{center}
\includegraphics[width=1.0\linewidth]{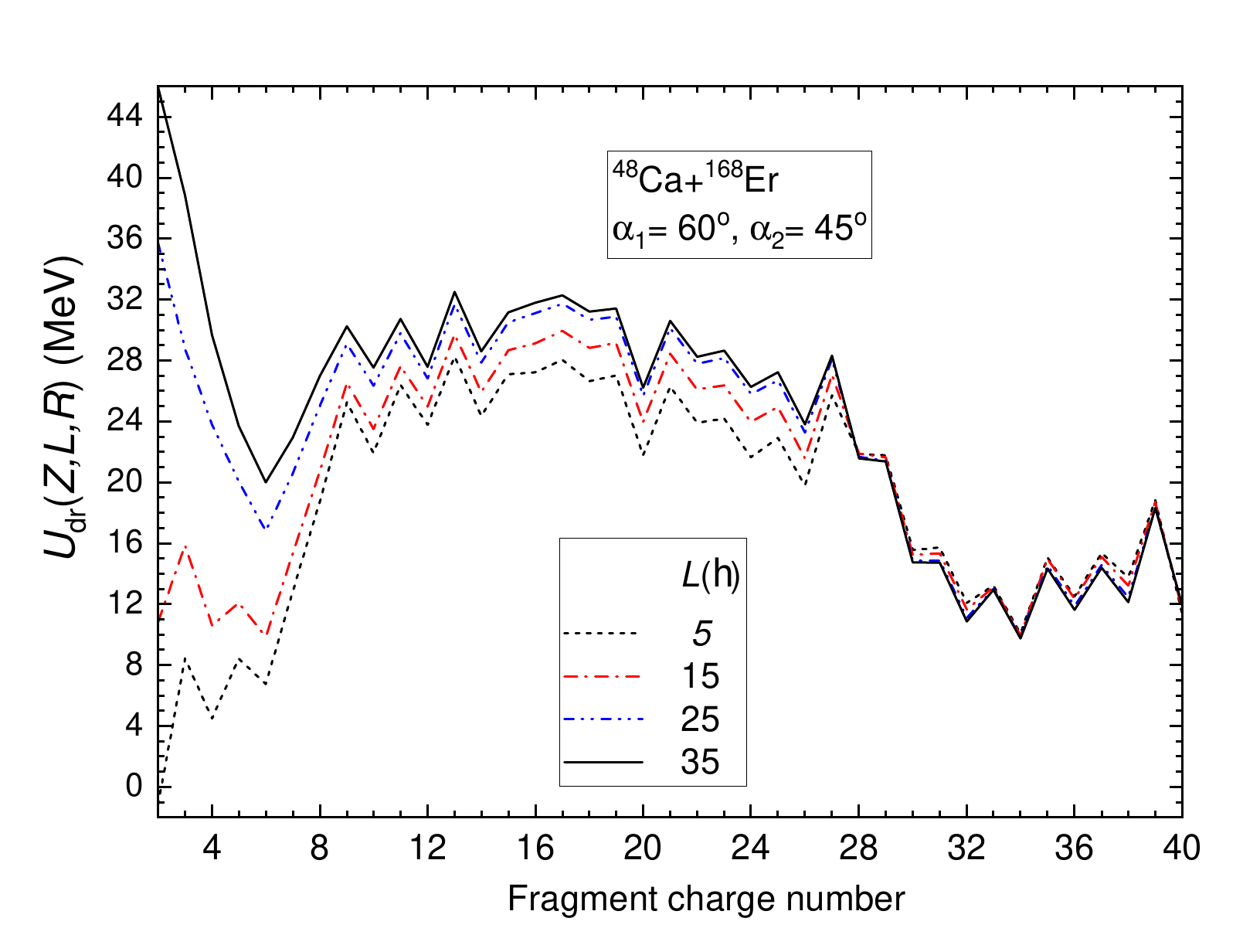}
  \vspace*{-0.35cm}
  \caption{The dependence of the driving potential on the angular momentum $L$. The results 
  are obtained for the collision with the orientation angles $\alpha_1=60^{\circ}$ 
  and $\alpha_2=45^{\circ}$.}
   \label{UdrL}
		\end{center}
\end{figure}
The PES $U$ calculated for the $^{48}$Ca+$^{168}$Er reaction,  driving potential $U_{dr}$ and
nucleus-nucleus interaction $V$ extracted from the PES are presented in Fig. \ref{PES2barr}. 
The arrow (a) corresponds to the capture trajectory and arrow (b) shows the direction to the complete fusion. The arrows (c) and (d) correspond to the possible quasifission trajectories. After capture, the DNS can follow to the CN formation along charge asymmetry axis $Z$ in the direction of its decreasing $Z\rightarrow 0$ or breakup channel along line $R$ connecting  centres-of-fragments. The minimum values of the PES along the charge asymmetry axis are observed when the proton and/or neutron numbers in the DNS fragments close to the magic numbers. The position of the entrance channel for the $^{12}$C+$^{204}$Pb reaction is favorable to complete fusion since the intrinsic barrier causing hindrance is very small while it is sufficiently larger the  $^{48}$Ca+$^{168}$Er reaction.  Fig. \ref{PES2barr}(b) is 
presented to show the determination of the intrinsic fusion barrier $B^*_{\rm fus}$  from the 
curve of the driving potential as difference between values of the  driving potential corresponding to $Z=20$ and its maximum value in direction to  complete fusion. 
The dependence of the driving potential on the angular momentum is presented in Fig. 
\ref{UdrL}.  The increase of the angular momentum leads to the increase of the $B_{\rm fus}^*$  up the large values for the very charge asymmetric configurations 
(for small values of $Z$) of DNS. This phenomenon is explained by the strong increase 
of the DNS rotational energy.  

\begin{figure}[ht]
		\begin{center}
			\includegraphics[width=1.0\linewidth]{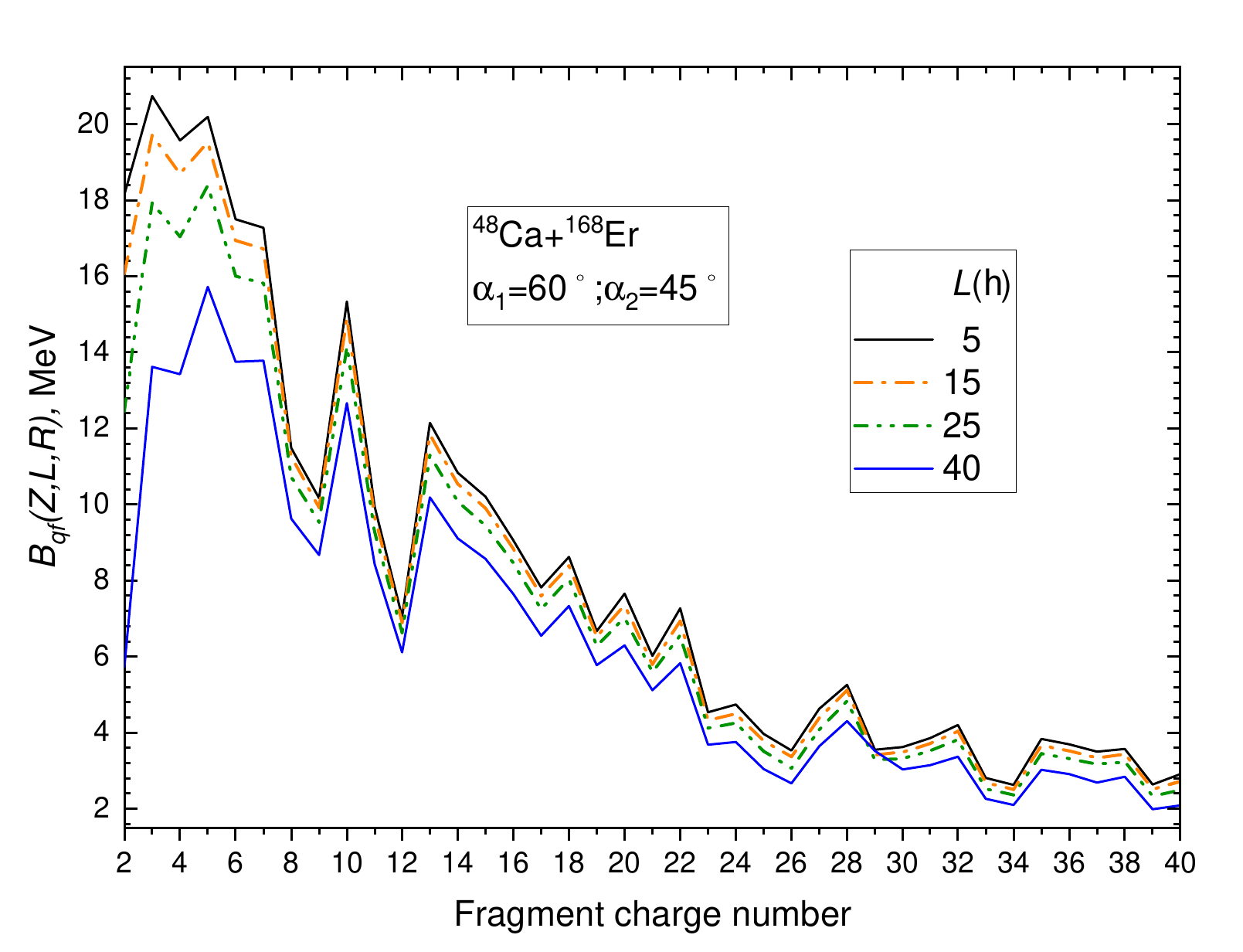} 
     \vspace*{-0.35cm}
  \caption{The quasifission barriers $B_{qf}$ calculated for the DNS fragments which can be formed in the $^{12}$C+$^{204}$Pb and $^{48}$Ca+$^{168}$Gd reactions in collisions with the orbital angular momentum $L$. The results   are obtained for the collision with the 
  orientation angles $\alpha_1=60^{\circ}$   and $\alpha_2=45^{\circ}$.}
     \label{Bqf}
		\end{center}
	\end{figure}
The rotational energy of the DNS with the light fragment having small mass numbers increases  faster due to the strong decrease of moment of inertia: $J_{DNS}(Z,A)=\mu R_m^2+(J_1(Z,A)+J_2(Z_c,A_c))/2$. It is used in calculation of the rotational energy:
\begin{eqnarray}
	V_{rot}(Z,A, L,R)=\frac{L(L+1)}{2J_{DNS}(Z,A,Z_c,A_c)}, 
 \label{Vrot}
\end{eqnarray}
where $\mu=mAA_c/(A+A_c)$ - reduced mass of DNS; $J_1=mA(a_1^2+b_1^2)/5$ and $J_2=mA_c(a_2^2+b_2^2)/5$ are moments inertia of the interacting nuclei; $m$ is a nucleon mass; $a_i$ and $b_i$ are small and large radii of nuclei \cite{Nasirov2023PLB}.

The fast increase of the rotational energy of the DNS with the light fragments is responsible for  the incomplete fusion in the reactions with light nuclei \cite{Nasirov2023PLB}.
  Therefore, in both reactions, when $L$ increases, the probability of fusion decreases. Another important physical quantity of the model is quasifission barrier $B_{qf}$ (see Fig. \ref{Bqf}) which determines the DNS lifetime.  Its value is equal to the depth of  the potential well of the nucleus-nucleus interaction between the DNS fragments. The height of fusion barrier $B_{\rm fus}^*$ for the reaction is less than the height of quasifission barrier $B_{qf}$ for the mass asymmetric system. This condition is favorable for the 
  complete fusion. 
  It is seen from Fig.  \ref{Bqf} that its value for the $^{12}$C+$^{204}$Pb ($Z=6$) system is sufficiently larger than the one for the $^{48}$Ca+$^{168}$Er ($Z=20$)  reaction.  For the last  reaction, the height of the fusion barrier $B_{\rm fus}^*$ is greater than the height of the quasifission barrier $B_{qf}$. Therefore, the  probability of complete fusion for the reaction  $^{12}$C+$^{204}$Pb is larger than one for the  $^{48}$Ca+$^{168}$Er 
  reaction. 

\begin{figure}[ht]
 \vspace*{-0.25cm}
\includegraphics[width=1.12\linewidth]{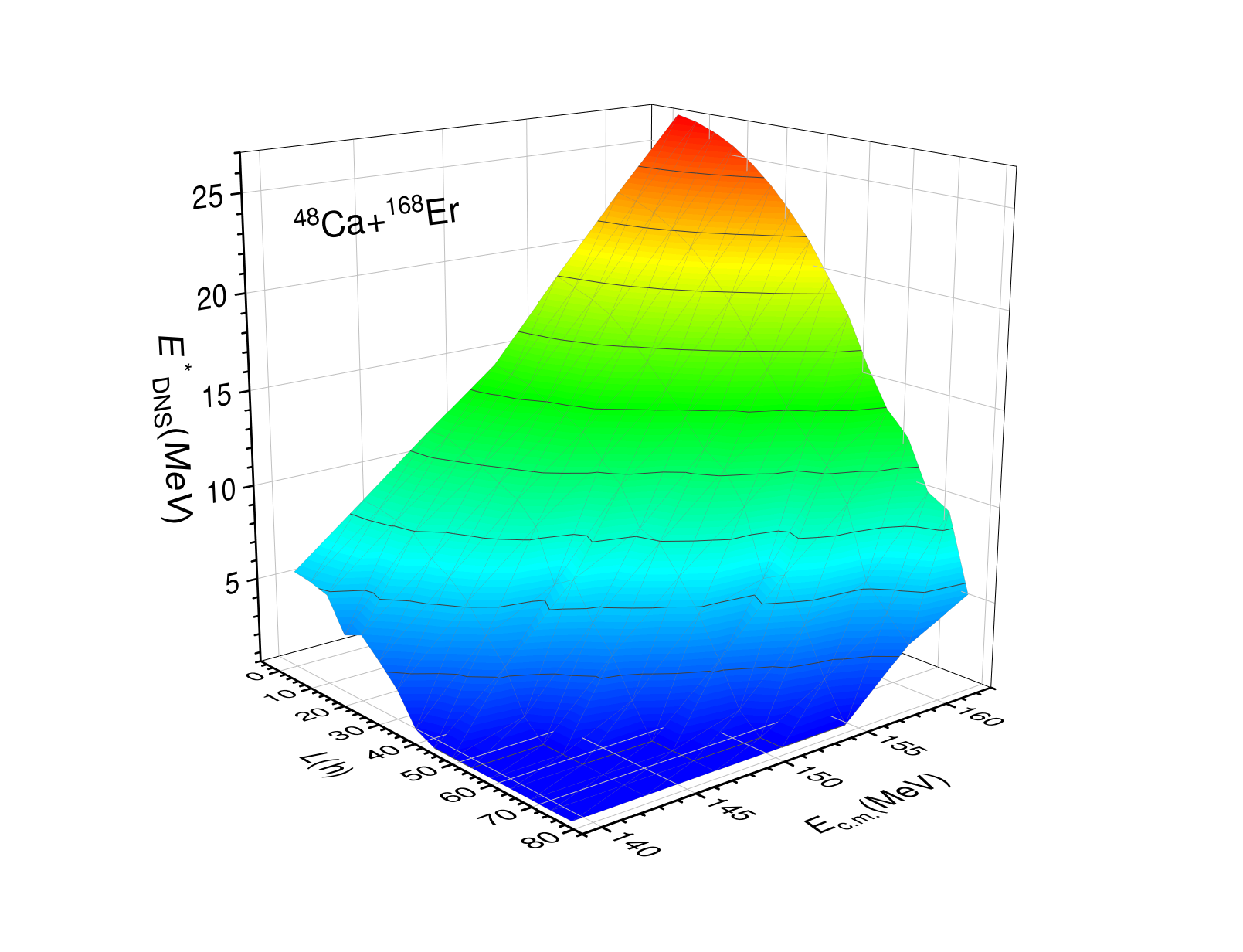}
  \vspace*{-0.55cm}
  \caption{The excitation energy $E_{\rm DNS}^*(E_{c.m.}, L)$ of the DNS formed in the  
   $^{48}$Ca+$^{168}$Gd reactions for  the entrance channel ($Z=Z_P$ and $A=A_P$) 
   as a function of the center mass energy $E_{\rm c.m.}$ and orbital angular momentum $L$.}
	\label{Edns}
\end{figure}

 The excitation energy of DNS $E_Z^*$, given the beam energy, is found taking into account the change in the internal energy with the change in the number of nucleons:
\begin{eqnarray}
E_Z^*(E_{\rm c.m.},A,L,\{\alpha_i\})&=&E_{\rm c.m.}+\Delta Q_{}gg(Z,A)\cr &-&
 V_{\rm min}(Z,A,R_m,L,\{\alpha_i\}),
 \end{eqnarray}
 where 
 \begin{eqnarray}
\Delta Q_{}gg(Z,A)&=&B_1(Z,A)+B_2(Z_c,A_c) \cr
                  &-&(B_P(Z_P,A_P)+B_T(Z_T,A_T))
\end{eqnarray}
is a change of the intrinsic energy of the DNS during its evolution from the initial value ($Z=Z_P$ and $A=A_P$) to the considered state $(Z,A)$.
 $V_{\rm min}(Z,A,R_m,L,\{\alpha_i\})$ is the minimum value of potential well of the nucleus-nucleus potential between the DNS  fragments in the last state \cite{Nasirov2005,Fazio2003}. 
 Fig. \ref{Edns} represents a dependence of the excitation energy of DNS $E_{\rm DNS}^*$ 
 for the entrance channel ($Z=Z_P$ and $A=A_P$) on the
 collision energy $E_{\rm c.m.}$ and its angular momentum $L$ calculated for the 
 $^{48}$Ca+$^{168}$Er reaction.   

\subsection{Charge and mass distribution of the DNS fragments and binary yields}

The full momentum transfer takes place at the capture of the projectile by the target nucleus 
and the DNS is formed with the probability $\mathcal{P}_{cap}$, which  is calculated by the solution of the dynamical equation of the collision trajectory for the given values of  $E_{\rm c.m.}$ and orbital angular momentum $L$ \cite{Nasirov2005,Fazio2003}. The atomic nucleus consists of neutrons and protons, consequently, due to the nucleon exchange between the DNS nuclei, their mass and charge distributions change as a function of time $t$ as capture has occurred.  
\begin{figure*}[ht!]
	\begin{minipage}[ht!]{0.475\linewidth}
		\begin{center}
			\includegraphics[width=1\linewidth]{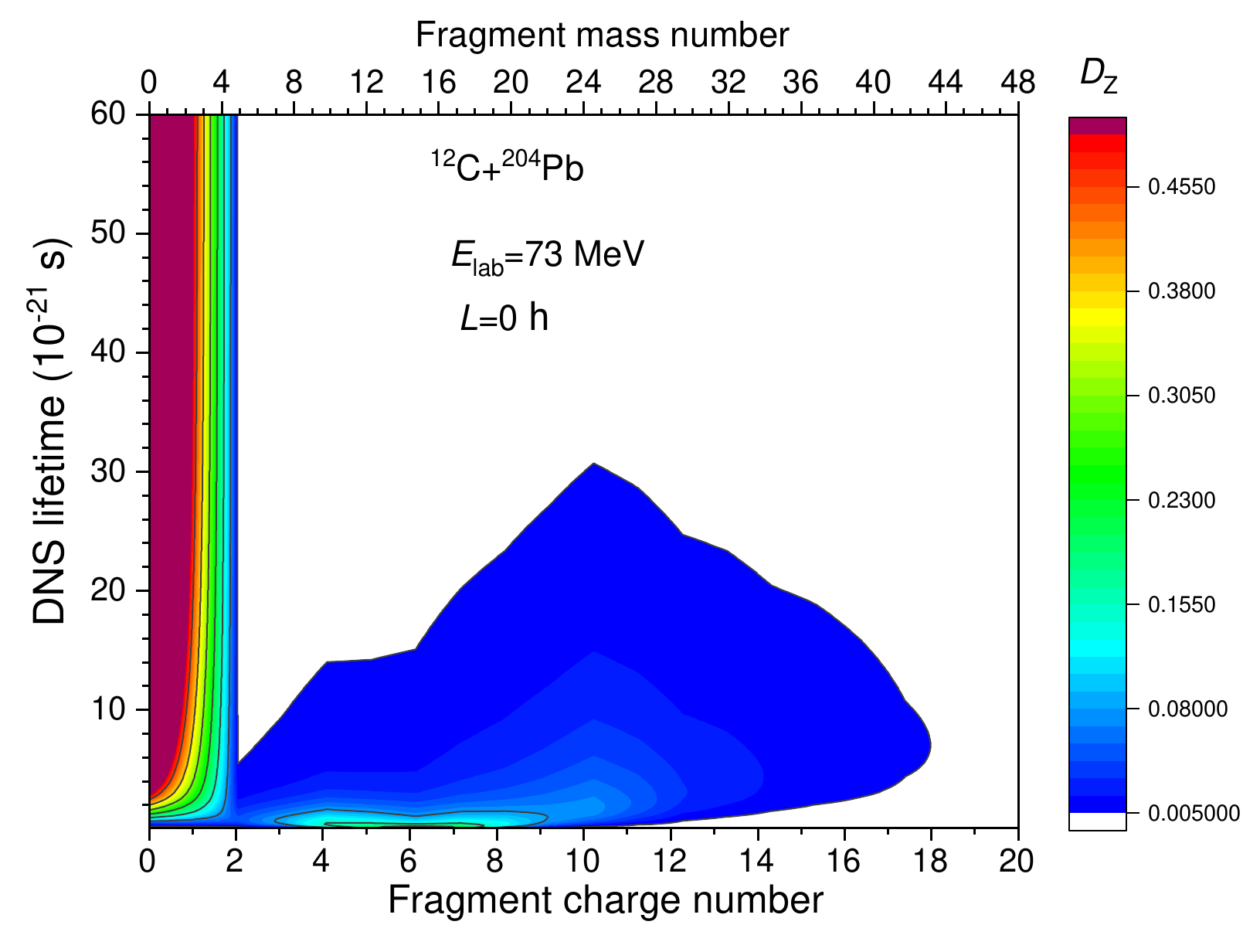}
		\end{center}
       \vspace{-0.65 cm}
  \caption{Evolution of the charge distribution $D_Z$ for the projectile-like fragments  for the	$^{12}$C+$^{204}$Pb reaction at $E_{\rm Lab}$ = 73 MeV and $L$ =0$\hbar$. The results have been obtained for the orientation angles $\alpha_1$ = 45\textdegree and $\alpha_2$ = 30\textdegree.}
 \label{DZ12CL0}
	\end{minipage}
	\hfill
	\begin{minipage}[ht!]{0.475\linewidth}
\vspace{-1.20 cm}
		\begin{center}
			\includegraphics[width=0.975\linewidth]{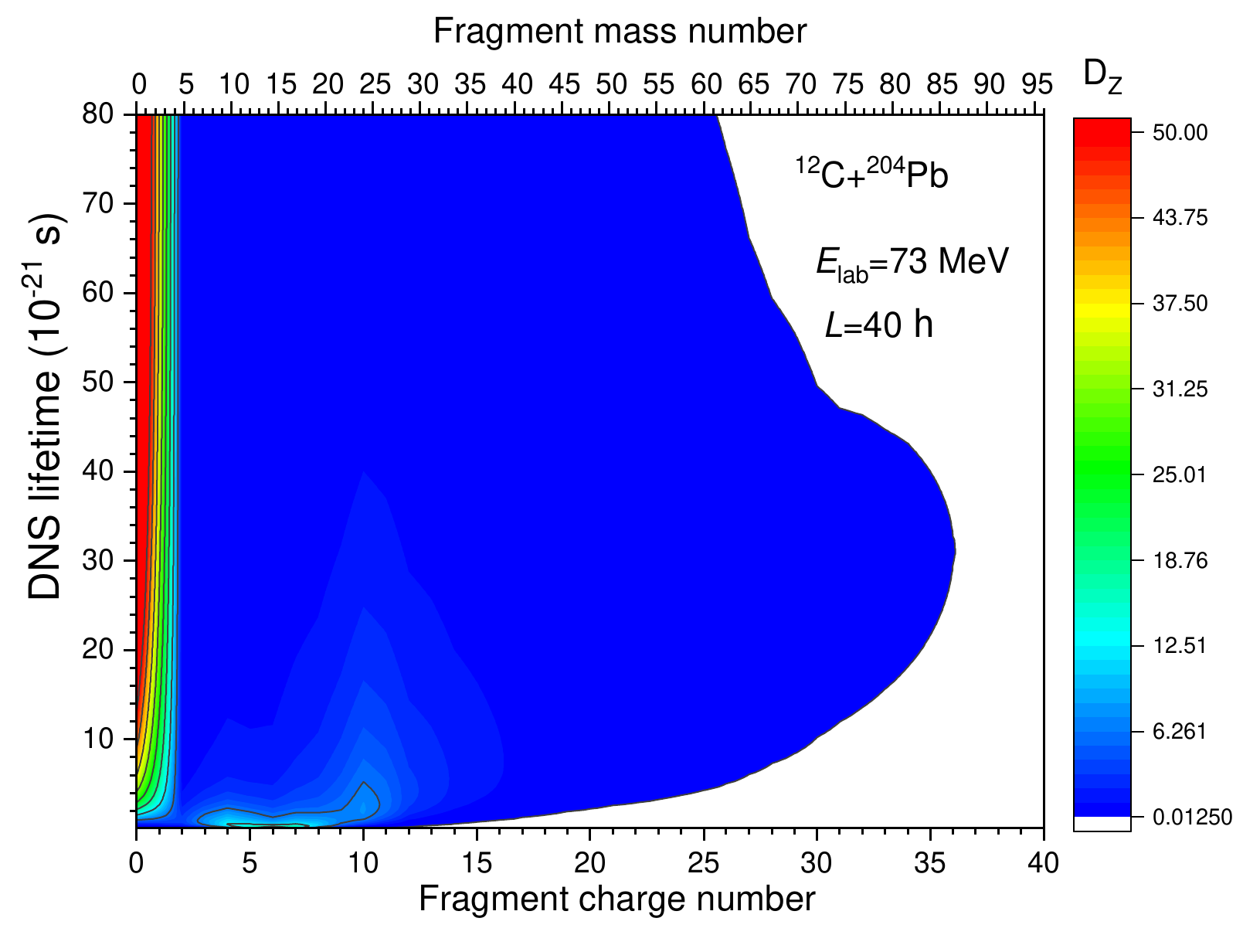}
		\end{center}
\vspace{-0.5 cm}
\caption{The same as in Fig. \ref{DZ12CL0}, but for    
 $L$ =40$\hbar$. }
	\label{DZ12CL40}
	\end{minipage}
\end{figure*}
The evolution of DNS is found by solving the transport master equation \cite{Nasirov2023PLB}:
\begin{eqnarray}
	\frac{\partial D_Z(t)}{\partial t}&=&\Delta_{Z+1}^{(-)}D_{Z+1}(t)+\Delta_{Z-1}^{(+)}D_{Z-1}(t)\cr &-&(\Delta_{Z}^{(-)}+\Delta_{Z}^{(+)})D_Z(t),\label{master}
\end{eqnarray}
where $D_{Z}(t)$ is represents the probability of DNS being in the moment of time $t$ for the
given $E_Z^*$ and $L$ in the state where the DNS fragments have the charge numbers 
$Z$ and $Z_{\rm CN}-Z$. $\Delta_{Z}^{\pm}$ ($\Delta_{Z}^{\pm}$) coefficients are the 
transport coefficients, which are calculated microscopically, for the case when one proton is 
added to (subtracted from) the fragment of the binary system with the charge number $Z$. 
Proton and neutron systems of nuclei have their own energy schemes in individual nuclei. In turn, these schemes depend on the number of neutrons $N$ and the number of charges $Z$ of the nuclei. This means that the quantities $\Delta_{Z}$ are related to the energy schemes of the protons (they fill the energy states).  We can calculate transport coefficients using the following formula:
\begin{eqnarray}
	&&\Delta_Z^{(+)}=\frac{1}{\Delta t}\sum_{P,T}|g_{PT}( R)|^2n_T^Z(\theta_T)(1-n_P^Z(\theta_P))\times\cr &&\dfrac{\sin^2\left(\Delta t[\varepsilon_P^Z-\varepsilon_T^Z]\right)/(2\hbar)}{(\varepsilon_P^Z-\varepsilon_T^Z)^2/4}\cr&&
	\Delta_Z^{(-)}=\frac{1}{\Delta t}\sum_{P,T}|g_{PT}( R)|^2n_P^Z(\theta_P)(1-n_T^Z(\theta_T))\times\cr&&\dfrac{\sin^2\left(\Delta t [\varepsilon_P^Z-\varepsilon_T^Z]\right)/(2\hbar)}{(\varepsilon_P^Z-\varepsilon_T^Z)^2/4}.
 \label{DeltaZ}
\end{eqnarray}
Here the matrix elements $g_{PT}$ represent the exchange of nucleons between fragments ``P'' 
and ``T''; $n_i^Z(T)$ and $\varepsilon_i^Z$ are occupation number and energy of 
single-particle states in fragment ``i'' of the DNS with a fragment with the charge 
number $Z$, respectively, $\theta_i$ is its 
effective temperature ( $i=P,T$):
\begin{equation}
   \theta_i= \sqrt{\frac{E^*_Z}{a}\left(\frac{A}{A_{\rm CN}}+\frac{1}{2}\right)}, 
\end{equation}
where $a=A_{\rm CN}/12$ MeV$^{-1}$.
The transport master equation is solved where the reaction time $t=k_{\rm max}\Delta t$, where $\Delta t=10^{-22}~s$. In this case, $t$ is chosen in such a way that after this moment of time, DNS has passed to complete fusion or quasifission.

The region $Z\geq 2$ represents the contribution of $D_Z$ to the incomplete fusion. We can calculate the yield of fragments formed in the reaction using the formula:
\begin{eqnarray}
Y_Z(E_Z^*,A,L,t)&=&\Lambda_Z^{\rm qf}(B_{\rm qf}(Z,A,\{\alpha_i\}),T_Z(A,\alpha_i))\cr &\times&\sum\limits_{k=0}^{k_{\rm max}}D_Z(A,E_Z^*,L,k\Delta t)\label{Yz}
\end{eqnarray}
It is proportional to the width $\Lambda_Z^{\rm qf}$ of the decay through the quasifission barrier $B_{\rm qf}(Z,A,L)$. $\Lambda_Z^{\rm qf}$ is calculated by expression:
\begin{eqnarray}
\Lambda_Z^{\rm qf}(Z,A,\{\alpha_i\},T_Z(A,\alpha_i))&=&\exp\left({\frac{-B_{\rm qf}(Z,A,\{\alpha_i\})}{T_Z(A,\alpha_i)}}\right) \cr &\times&\frac{\Gamma_Z\omega_m(Z)}{2\pi\omega_{\rm qf}(Z)}
\end{eqnarray}
where $\omega_m$ and $\omega_{\rm qf}$ are frequencies of the parabolas used to approximate the potential well and Coulomb barrier of the nucleus-nucleus interaction; $\gamma=8\cdot10^{-22}$~MeV~sec$^{-1}$; $T_Z$ is the effective temperature of the DNS with the charge asymmetry $Z$: $T_Z(A,\alpha_i)=\sqrt{E_Z^*(A,L,\alpha_i)/a},~a=A_{\rm tot}/12$
MeV$^{-1},~A_{\rm tot}=A_P+A_T$; 
\begin{eqnarray}
	\Gamma_Z=\sqrt{\frac{\gamma^2}{(2\mu)^2}+\omega_{\rm qf}^2(Z)}-\frac{\gamma}{2\mu}.
\end{eqnarray}
(see Refs \cite{NasirovNPA2016, Kayumov2022} for details).

Eq. (\ref{master}) has been solved with the initial condition $D_Z$=1 at $Z=Z_P(Z_T)$ and $A=A_P(A_T )$. 
The charge distributions $D_Z$ for the light fragment of the DNS formed in the $^{12}$C+$^{204}$Pb reaction in collisions with the values of  $L$=0$\hbar$ and $L$=40$\hbar$ at the beam energy $E_{\rm Lab}=73$MeV  are presented in Figs. \ref{DZ12CL0}  and \ref{DZ12CL40}, respectively. The presented results have been obtained for the orientation angles of ``P'' and ``T'' nuclei $\alpha_1$ = 45\textdegree and $\alpha_2$ = 30\textdegree, respectively. It is seen that the intense of the nucleon transfer from $^{12}$C to $^{204}$Pb decreases by the increase of $L$. The yields with  $Z<2$,  represents the contribution leading to complete fusion. The complete fusion occurs faster since the charge distribution ($D_Z$) in the region $Z>2$ is very weak. 

 It is seen from Figs. \ref{DZ12CL0} and \ref{DZ12CL40} that at the beginning the charge distribution  is distributed around $Z_P=6$ in the light fragment (for the conjugate fragment around  $Z_T=82$, it is not  shown). The fusion  barrier $B_{\rm fus}^*(Z=6)$ is small for the  entrance channel of the $^{12}$C+$^{204}$Pb reaction (see Fig.  \ref{PES}   of the PES and 
 driving potential $U_{\rm dr}$, respectively). Moreover, the quasifission barrier $B_{\rm qf}$ is large for the charge asymmetric configurations of DNS (see Fig. \ref{Bqf}). This favorable conditions cause the motion of the charge distribution towards $Z=2$  over time. Consequently,  the complete fusion takes place with the larger probability than quasifission process for this strong charge asymmetric reaction.
\begin{figure*}[ht!]
	\begin{minipage}[ht!]{0.46\linewidth}
		\begin{center}
			\includegraphics[width=1\linewidth]{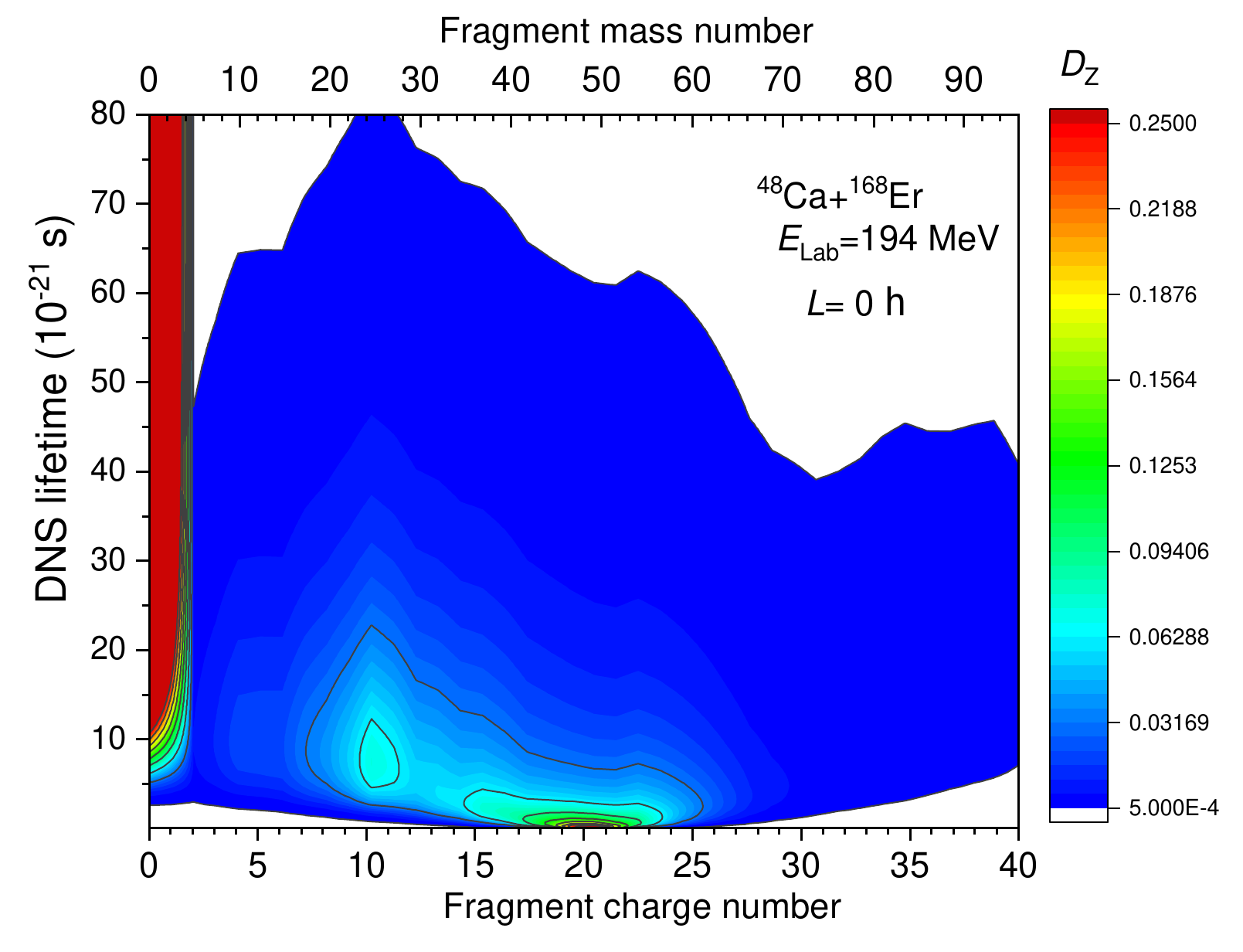} 	
		\end{center}
       \vspace{-0.65 cm}
  	\caption{Evolution of the charge distribution $D_Z$ for the projectile-like fragments for the $^{48}$Ca+$^{168}$Er reaction at $E_{\rm Lab}$ = 194 MeV and $L$ =0$\hbar$. The results have been obtained for the orientation angles $\alpha_1$ = 45\textdegree and $\alpha_2$ = 30\textdegree.}
  \label{Dz48CaL0}
	\end{minipage}
  	\begin{minipage}[ht]{0.46\linewidth}
   \vspace{-1.0 cm}
      \hspace{0.25 cm}
\includegraphics[width=1\linewidth]{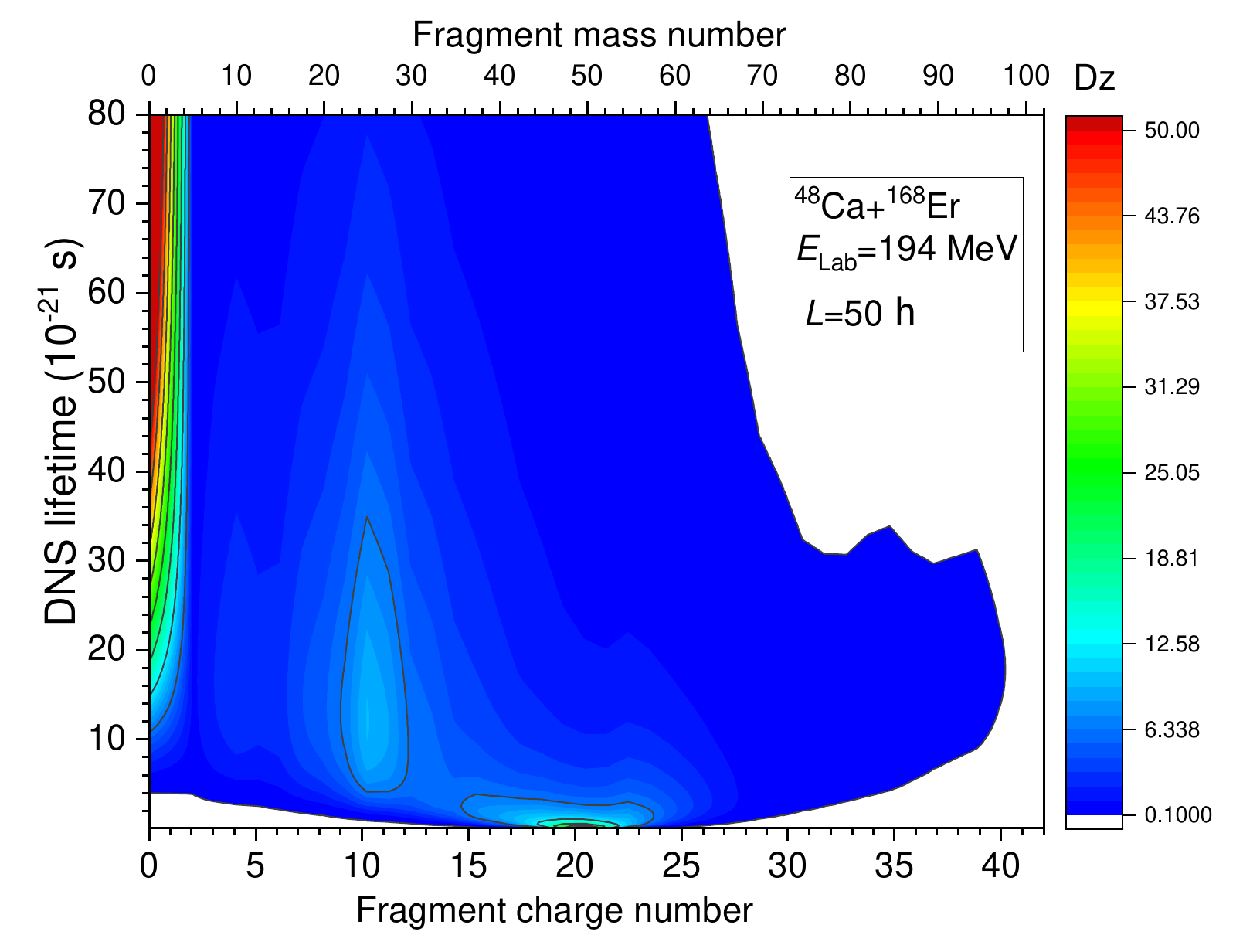}
   \vspace{-0.730 cm}
  	\caption{The same as in Fig. \ref{Dz48CaL0} but for  $L$ =50$\hbar$. 
 }
    \label{Dz48CaL50}
	\end{minipage}
\end{figure*}
 
Figs. \ref{Dz48CaL0} and \ref{Dz48CaL50} shows that, in the case of the $^{48}$Ca+$^{168}$Er reaction charge is concentrated distributed around $Z_P=20$  (for the heavy fragment around $Z_T=68$) at the beginning of the DNS evolution and it is distributed wider including the direction of the larger charge numbers $Z > 20$. The presence of the hindrance to complete fusion in the case of the $^{48}$Ca+$^{168}$Er reaction in comparison with the $^{12}$C+$^{204}$Pb reaction is clearly seen from these figures. Therefore,  the complete fusion  occur more slowly in  the $^{48}$Ca+$^{168}$Er reaction for the both values of $L$. The other reason of the observation is the fact that the quasifission barrier $B_{\rm qf}(Z=20)$ in the entrance channel of the $^{48}$Ca+$^{168}$Er reaction is smaller than the one for the $^{12}$C+$^{204}$Pb reaction (see Fig. \ref{Bqf}). The small values of  $B_{\rm qf}$ is favorable for the quasifission.

\begin{figure*}[ht!]
	\begin{minipage}[ht!]{0.48\linewidth}
		\begin{center}
			\includegraphics[width=1\linewidth]{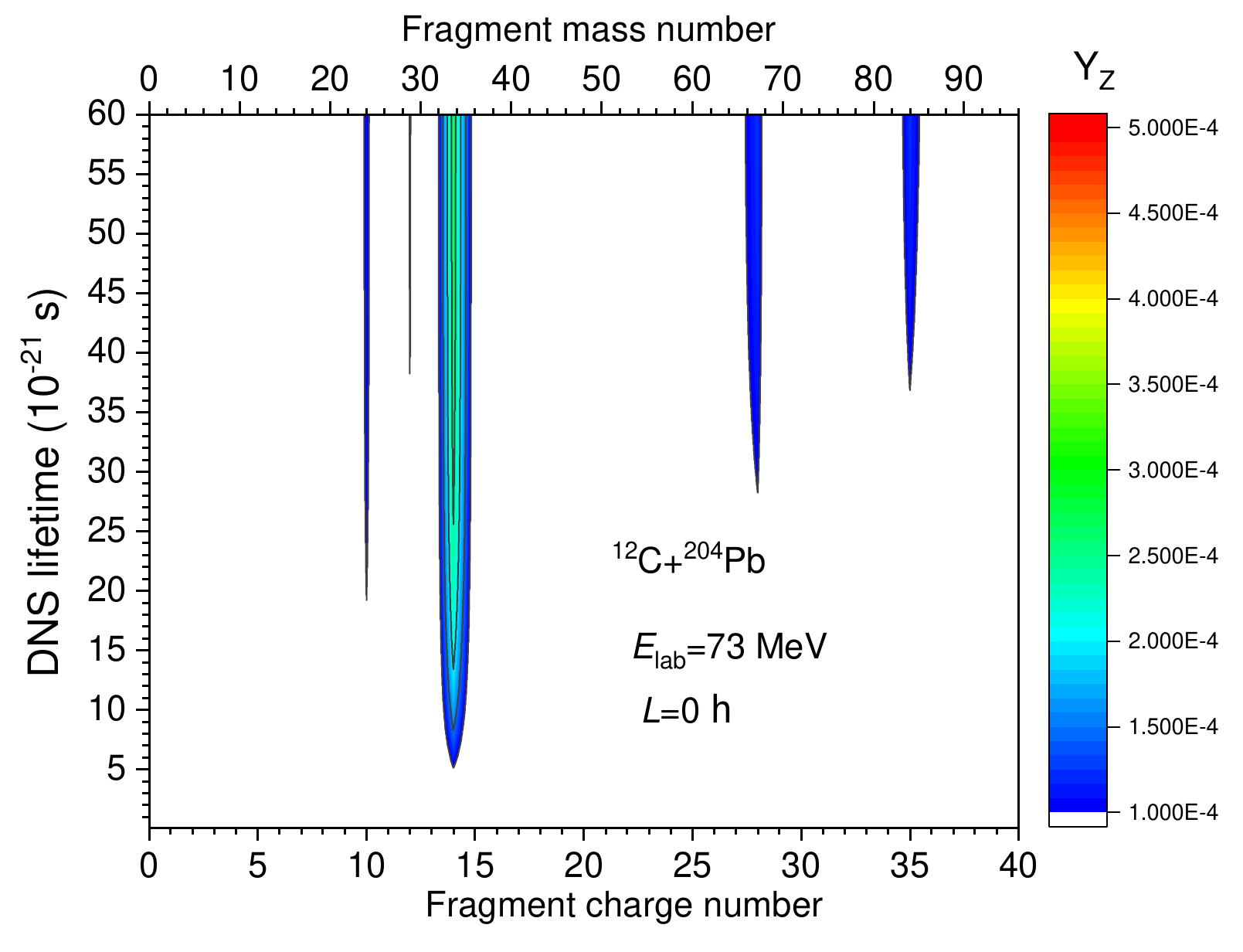}  
		\end{center}
\vspace{-0.65 cm}
  \caption{ Mass distribution of the yield of quasifission products ($Y_Z$) for the reaction $^{12}$C+$^{204}$Pb calculated for the collision with the orientation angles $\alpha_1$ = 45\textdegree and $\alpha_2$ = 30\textdegree at the beam energy $E_{\rm Lab}$=73 MeV and  angular momentum $L=0$.}
  \label{YzC12Pb204L0}
	\end{minipage}
	\hfill
	\begin{minipage}[ht!]{0.48\linewidth}
		\begin{center}
			\includegraphics[width=1\linewidth]{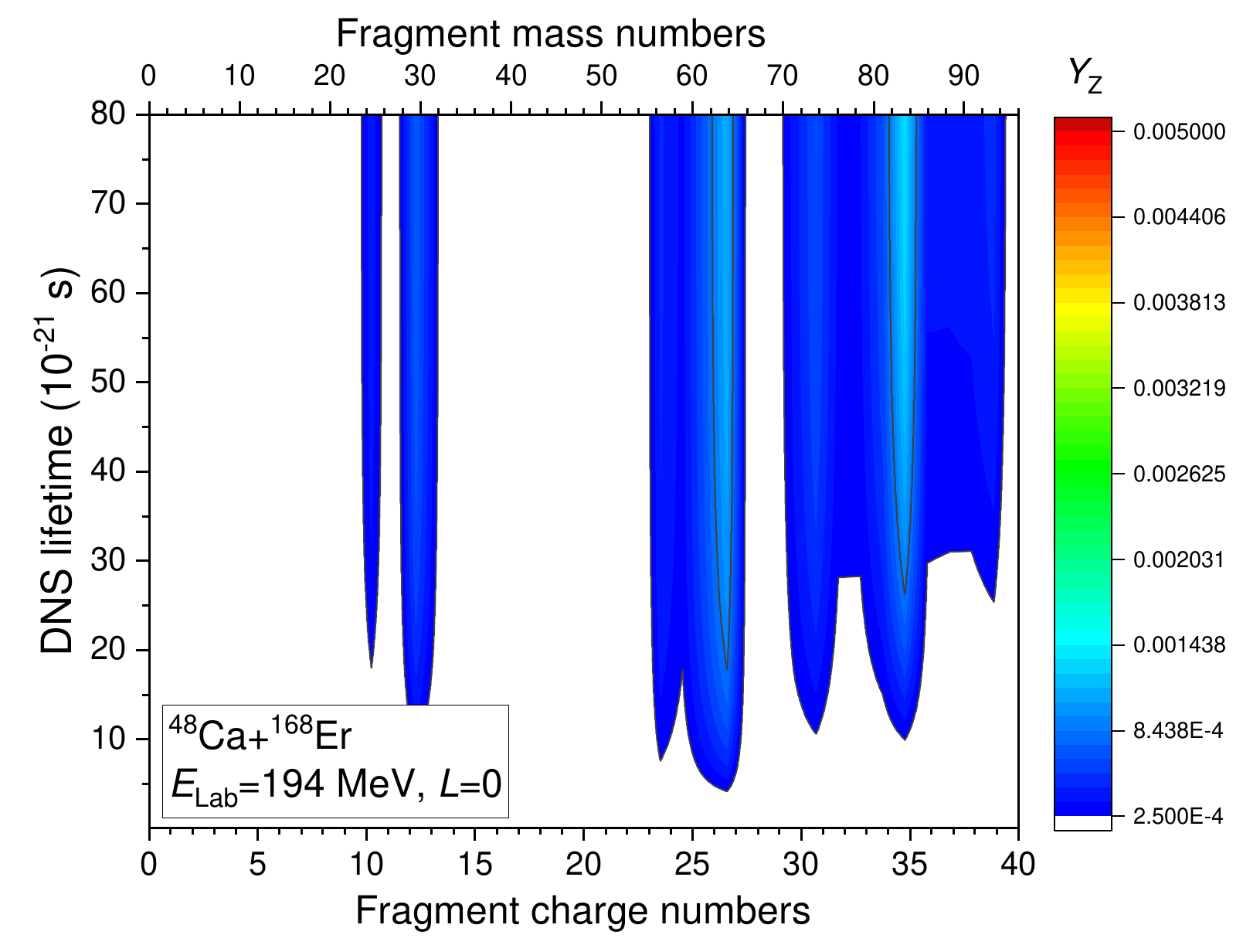} 
		\end{center}
\vspace{-0.65 cm}
  \caption{ Mass distribution of the yield of quasifission products ($Y_Z$) for the reaction $^{48}$Ca+$^{168}$Er calculated for the collision with the orientation angles $\alpha_1$ = 45\textdegree and $\alpha_2$ = 30\textdegree at the beam energy $E_{\rm Lab}$=194 MeV and  angular momentum $L=0$.}	
  \label{Yz48Ca168ErL0}
	\end{minipage}
\end{figure*}

The yield  $Y_Z$  depends on the DNS angular momentum $L$ which determines the heights of the intrinsic fusion $B^*_{\rm fus}$ and quasifission $B_{\rm qf}$ barriers. It can be seen from Fig.\ref{Bqf}  that the quasifission barrier is changed as a function of $L$: at large values of $L$ the DNS becomes less stable against breakup into two quasifission fragments. As it has been mentioned above that $B^*_{\rm fus}$ increases by $L$ (see Fig. \ref{PES}). Therefore, the quasifission yields increase by the increase of $L$.
 
\begin{figure*}[ht!]
	\begin{minipage}[ht!]{0.48\linewidth}
		\begin{center}
			\includegraphics[width=1\linewidth]{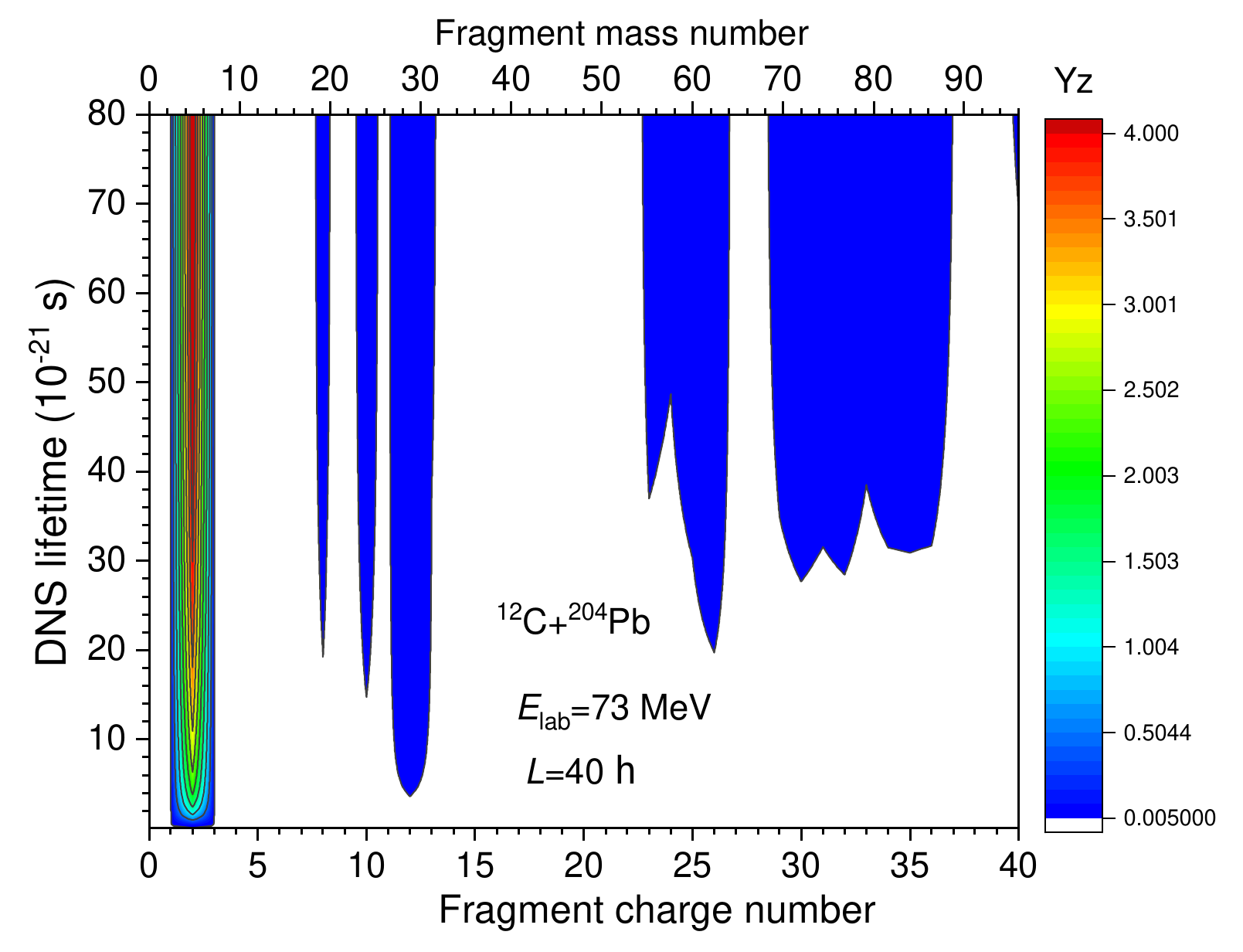} 
		\end{center}
  \vspace{-0.6 cm}
  	\caption{ Mass distribution of the yield of the quasifission products ($Y_Z$) in the 
  $^{12}$C+$^{204}$Pb  reaction  calculated for the  orbital angular momentum $L=40 \hbar$.
  The result has been  obtained for the orientation angles $\alpha_1$ = 45\textdegree and 
  $\alpha_2$ = 30\textdegree.}
  \label{YzL40C12Pb204}
	\end{minipage}
	\hfill
	\begin{minipage}[ht!]{0.48\linewidth}
		\begin{center}
			\includegraphics[width=1\linewidth]{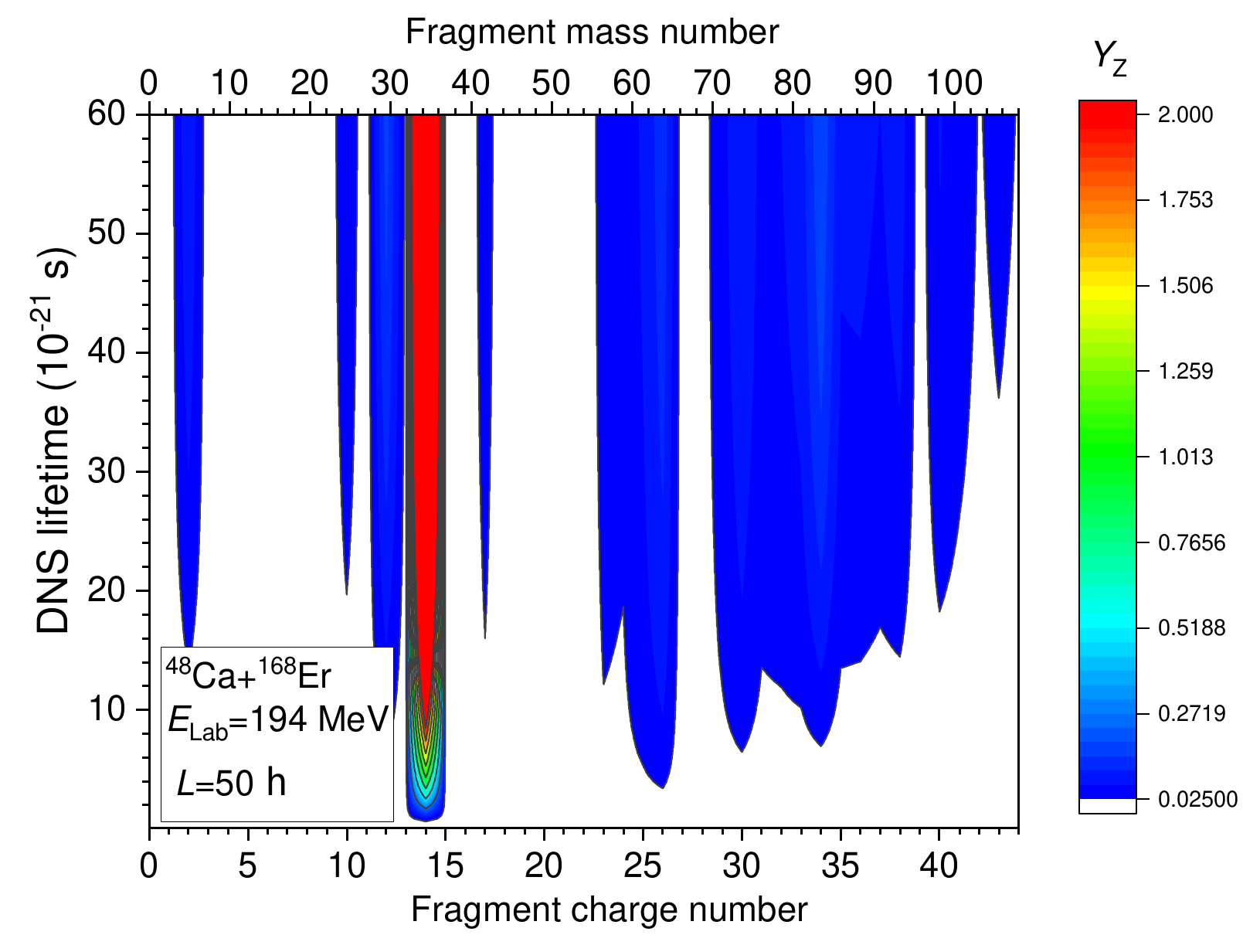} 
		\end{center}
  \vspace{-0.6 cm}
  \caption{ Mass distribution of the yield of the quasifission products ($Y_Z$) in the 
   $^{48}$Ca+$^{168}$Er  reaction  calculated for the  orbital angular momentum $L=50\hbar$.
      The results have been  obtained for the orientation angles $\alpha_1$ = 45\textdegree 
      and $\alpha_2$ = 30\textdegree.}
  \label{YzL5048Ca168Er}
	\end{minipage}
	\end{figure*}

Comparison of Figs. \ref{YzC12Pb204L0} and \ref{YzL40C12Pb204} of the calculated yields of binary fragments in the $^{12}$C+$^{204}$Pb reaction shows the strong increase of $Y_Z$ for the collisions with $L=40 \hbar$ in comparison with the case of $L=0 \hbar$. The similar increase of the binary yields is seen from the comparison of Figs. \ref{Yz48Ca168ErL0} and \ref{YzL5048Ca168Er} of the yields of the binary fragments calculated for the collisions $^{48}$Ca and $^{168}$Er with the orbital angular momentum $L=0$, respectively. The scales of $Y_Z$ presented on the right side of these figures show that the absolute values of the quasifission yields are small. This means that complete fusion is main reaction channel for the head on collision.  
The analysis of the yield products for  the $^{48}$Ca+$^{168}$Er reaction in Fig. \ref{YzL5048Ca168Er} and   $^{12}$C+$^{204}$Pb reaction in Fig.  \ref{YzL40C12Pb204}  shows that the main emitted products are $\alpha$ particles in collisions with the large angular momentum. This process is observed as the incomplete fusion according to its new mechanism verified in Ref. \cite{Nasirov2023PLB}. Therefore, the mechanism of the incomplete fusion 
can be considered as the yield of the very mass asymmetric quasifission products, {\it i. e.} 
the breakup of DNS  in the way to complete fusion due to increase of the centre-fugal forces 
at reaching $\alpha$ particle during multinucleon transfer from the projectile nucleus to the target nucleus.

\begin{figure*}[ht!]
	\begin{minipage}[ht!]{0.48\linewidth}
		\begin{center}
			\includegraphics[width=1\linewidth]{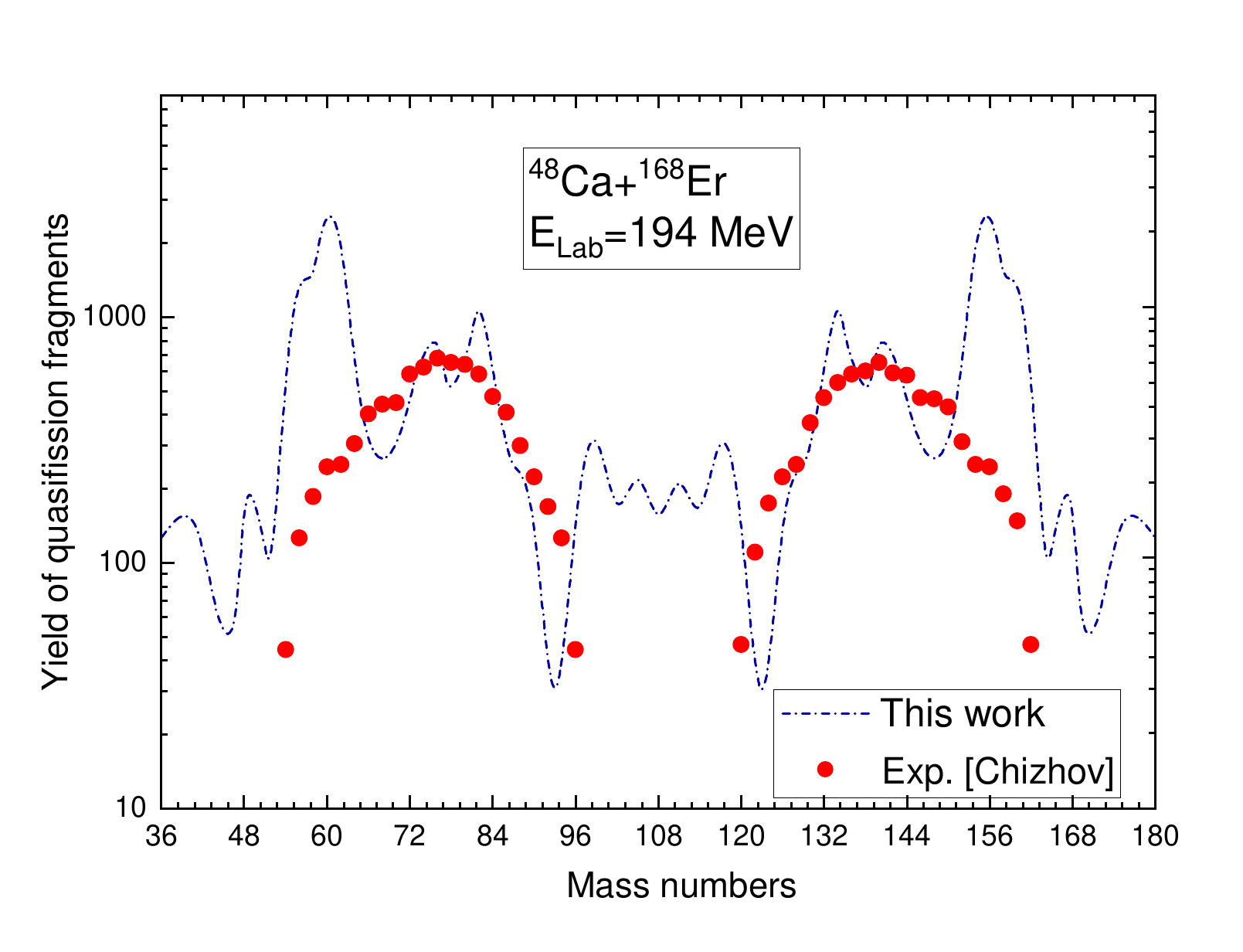} 
		\end{center}
    \vspace{-0.8 cm}
  	\caption{ Comparison of the theoretical yield of the quasifission products of the    $^{48}$Ca+$^{168}$Er  (at $E_{Lab}$=195 MeV)  reaction formed by the DNS mechanism with the corresponding measured experimental data obtained from Ref. \cite{Chizhov2003}. }
	\label{CompExpCa}
	\end{minipage}
	\hfill
	\begin{minipage}[ht!]{0.48\linewidth}
		\begin{center}
			\includegraphics[width=1\linewidth]{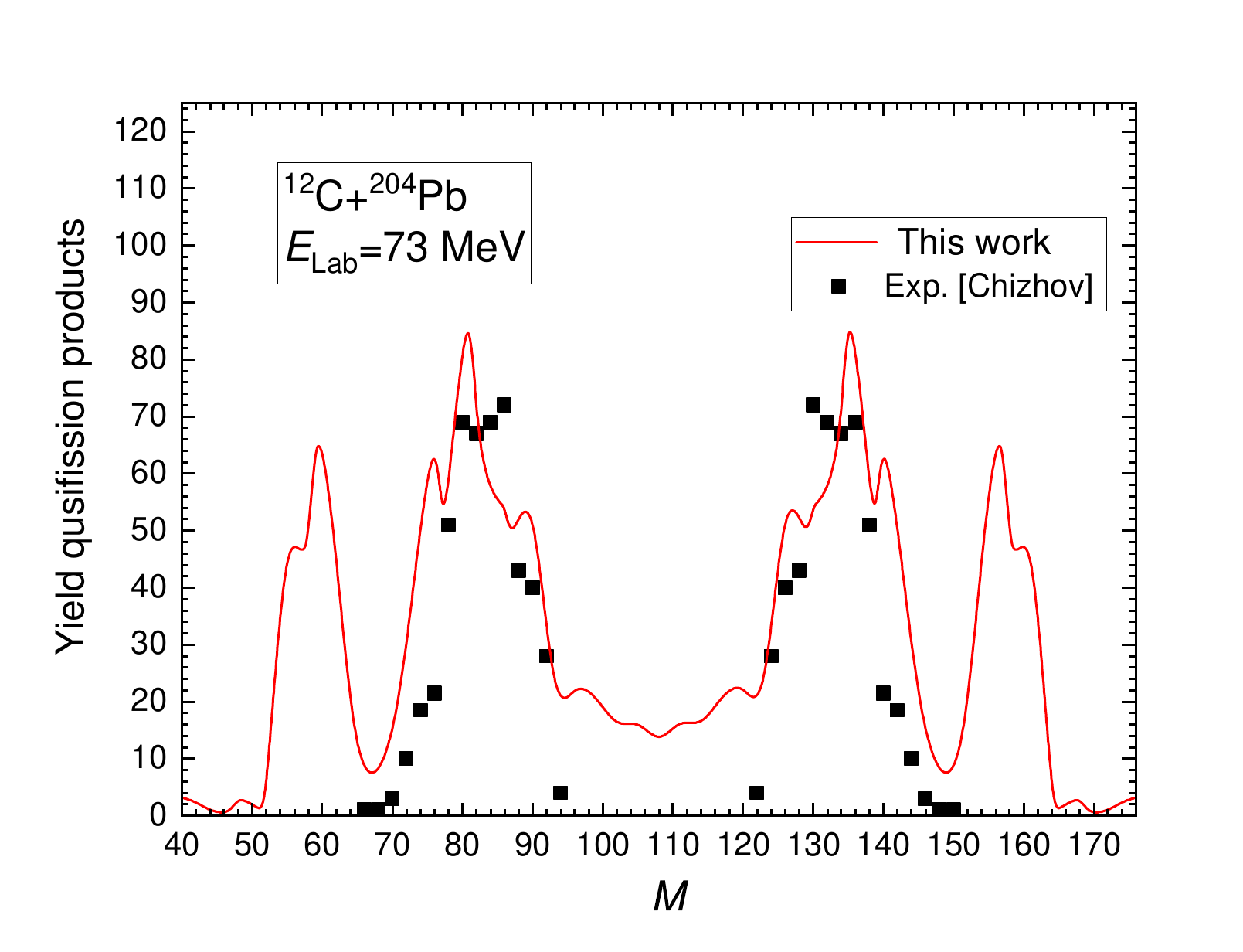} 
		\end{center}
  \vspace{-0.8 cm}
  	\caption{ Comparison of the theoretical yield of the quasifission products of the    $^{12}$C+$^{204}$Pb (at $E_{\rm Lab}$= 73 MeV) reaction formed by the DNS mechanism with the corresponding measured experimental data obtained from Ref. \cite{Chizhov2003}. }
	\label{CompExpC12}
	\end{minipage}

\end{figure*}
\section{Description of the experimental data}

Our calculations show that the charge distributions between fragments of the DNS and products being formed at its breakup strongly depend on the orbital angular momentum. The increase of the DNS rotational energy causes the increase in the intrinsic fusion barrier $B^*_{\rm fus}$, decrease of the  quasifission  barrier $B_{qf}$ and DNS excitation energy $E^*_Z$. These quantities and nuclear structure of the colliding nuclei  determines the intense nucleon exchange and direction flow of nucleons since the transition coefficients on the single-particle energies of the nucleons. Therefore, in this work, the shell effects of nuclear structure in fragments are pronounced in the formation and yield of the reaction products. These conclusions have been obtained from the dependence of the evolution of the charge distributions between fragments of the DNS in the $^{12}$C+$^{204}$Pb and $^{48}$Ca+$^{168}$Er reactions on the charge asymmetry and orbital angular momentum in the entrance channel of collision. 

It is important to prove the validity of this formalism of complete fusion by the description of the experimental data related with the yield of the quasifission products.

In  Figs. \ref{CompExpCa} and \ref{CompExpC12}, the mass distributions of the quasifission products the $^{48}$Ca+$^{168}$Er (at $E_{Lab}$=194 MeV)  and $^{12}$C+$^{204}$Pb (at $E_{\rm Lab}$= 73 MeV) reactions, respectively, calculated in this work  are compared with the corresponding experimental data obtained from Ref. \cite{Chizhov2003}. The experimental results are the extracted asymmetric components obtained as a difference 
\begin{equation}
    Y_{\rm qf}=Y_{\rm exp}-Y_{\rm FF},
    \label{symexp}
\end{equation}
where $Y_{\rm exp}$ is the experimental yield of the fission-like binary products and $Y_{\rm FF}$ is the (Gaussian) fusion-fission yield. The curve of $Y_{\rm FF}$ has been extracted by the description of the  experimental yield $Y_{\rm exp}$ as a sum of the three Gaussian functions: two small functions describing mass-asymmetric parts and one $Y_{\rm FF}$ which describes the mass-symmetric parts \cite{Chizhov2003}. This way separation of the quasifission and fusion-fission products assumes that there is not overlap between quasifission and fusion-fission products in the mass symmetric region of the mass distribution of the binary fragments.

But according our calculations shows the mass distribution of the quasifission can reach mass symmetric region. Its contribution is very small to the mass symmetric  in the case of the very mass asymmetric $^{12}$C+$^{204}$Pb reaction  even at large values of $L$. The yield of the quasifission products with the mass numbers $A>96$ is significant. It is seen Fig. \ref{CompExpCa}. The theoretical results are in good agreement with the experimental data for the range of mass numbers $A=$66--96 (120--150). The symmetric part  $A=$97--119 of the mass distribution of the binary products in the experimental data has been removed by Eq. (\ref{symexp}) while the curve of the theoretical results shows that the quasifission contribution presents in the mass symmetric region. 

\section{Conclusions}

In conclusion, we have theoretically studied charge and mass distributions of the 
quasifission fragments for two —
$^{12}$C+$^{204}$Pb and $^{48}$Ca+$^{168}$Er reactions — that lead to the same compound 
nucleus $^{216}$Ra$^*$ as a function of the orbital momentum of collisions at the 
beam energies corresponding to the CN excitation energy of around 40 MeV. 
The experimentally observed yield of the asymmetric fission in the former reaction 
was 1.5\%, whereas it was 30\% in the latter case. This difference was interpret 
as a large contribution of the quasifission products in the $^{48}$Ca+$^{168}$Er reaction.
Application of the DNS model has allowed us to establish a nature of hindrance to complete 
fusion by comparison results of the partial capture cross sections,   
charge ($D_Z$) and mass distributions of the DNS fragments before its breakup and 
quasifission ($Y_Z$) products obtained for the $^{12}$C+$^{204}$Pb and $^{48}$Ca+$^{168}$Er 
reactions. The theoretical study of the evolution of the charge ($D_Z$)  distributions of DNS 
fragments and quasifission ($Y_Z$) products shows strong influence of the 
orbital angular momentum of collision ($L$) on the hindrance of the complete fusion 
process. The difference in the hindrance  observed in these reactions is related 
by the intrinsic fusion barrier $B^*_{\rm fus}$ determined by the driving potential
calculated for the reactions leading to formation of the compound nucleus $^{216}$Ra$^*$.

The partial capture cross sections calculated for the $^{12}$C+$^{204}$Pb and $^{48}$Ca+$^{168}$Er reactions are very different and their maximum values 
are close to the critical angular momentum values presented in Ref. \cite{Chizhov2003}. 

The comparisons of the partial capture cross sections and charge (mass) 
distributions of the quasifission fragments calculated in this work for these two 
$^{12}$C+$^{204}$Pb and $^{48}$Ca+$^{168}$Er reactions show that the role of the 
entrance channel characteristics is very strong. This result confirms the 
conclusion of the authors of Ref. \cite{Chizhov2003}.

\appendix
\section*{Appendix}
\begin{figure}[ht!]
\centering
\includegraphics[scale=0.44]{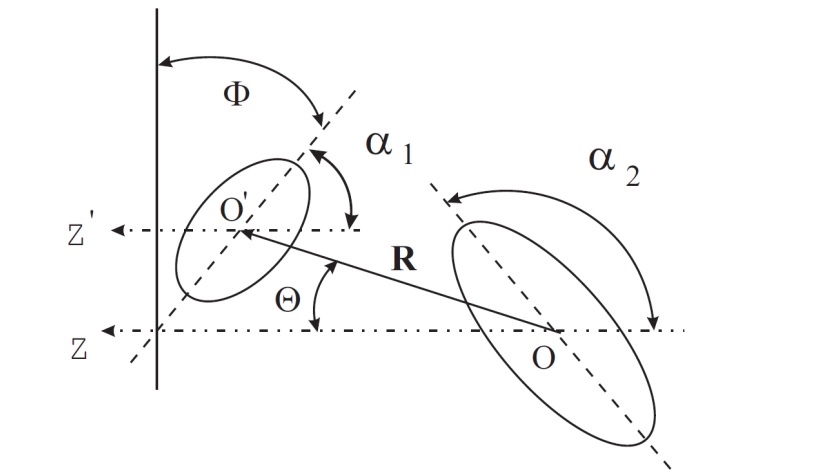}
\caption{The coordinate systems and angles which were used for the description of the initial orientations of projectile and target nuclei. The beam direction is opposite to $OZ$}\label{Burchak}
\end{figure}

The Coulomb interaction of deformed nuclei can be calculated according to the following expression:
\begin{eqnarray}
    &&V_{Coul}(Z_i,A_i,l,R,\{\alpha_i,\beta_i\})=\frac{Z_1Z_2}{R}e^2+\frac{Z_1Z_2}{R^3}e^2\times\cr &&\lbrace \left(\frac{9}{20\pi}\right)^{1/2}\sum_{i=1}^2  R_{0i}^2\beta_2^{(i)}P_2(\cos\alpha_i')+\cr &&\frac{3}{7\pi}\sum_{i=1}^2  R_{0i}^2\left[\beta_2^{(i)}P_2(\cos\alpha_i')\right]^2 \rbrace
\end{eqnarray}
where $\alpha_1'=\alpha_1+\theta,~\alpha_2'=\pi-(\alpha_2+\theta),~\sin{\theta}=|L|/(\mu\dot R R);~Z_i,~\beta_2^{(i)}$ and $\alpha_i'$ are the atomic number ( for each fragment ), the quadrupole deformation parameter, and the angle between the line connecting the centers of masses of the nuclei ( see Fig. \ref{Burchak}) and the symmetry axis of the fragment $i(i=1,2)$, respectively. Here, $R_{0i}=r_0A_i^{1/3},~r_0=1.16$~fm ,$e^2=1.44$~MeV$\cdot$ fm and $P_2(\cos{\alpha_i'})$ is the second term of the second type of Legendre polynomial; $\mu=M_1M_2/(M_1+M_2)$ is reduced mass of the colliding system consisting from projectile and target with masses $M_2$ and $M_2$, respectively.

The nuclear part of the nucleus-nucleus potential is calculated using the folding procedure between the effective nucleon-nucleon forces $f_{eff}[\rho(x)]$ suggested by Migdal and the nucleon density of the projectile and target nuclei, $\rho_1^{(0)}$ and $\rho_2^{(0)}$, respectively:
\begin{equation}
V_{nucl}(R)=\int \rho_1^{(0)}(r-r_1)f_{eff}(\rho)\rho_2^{(0)}(r-r_2)d^3r,
\end{equation}
\begin{equation}
 f_{eff}(\rho)=C_0 \left(f_{in}+(f_{ex}-f_{in})\frac{\rho (0)-\rho(r)}{\rho (0)}\right) \label{eff}
\end{equation}
Here $C_0$=300 MeV$\cdot$fm$^3$, $f_{\rm in}=0.09,~f_{\rm ex}=-2.59$ are the constants of the effective nucleon-nucleon interaction; $\rho=\rho_1^{(0)}+\rho_2^{(0)}$ .The effective values of the constants $f_{in}$ and $f_{ex}$ were fixed from the description of the interaction of the Fermi system by the Green function method and, therefore, the effect of the exchange term of the nucleon-nucleon interactions was taken into account. The densities of the interacting nuclei taken in the Woods-Saxon form:
\begin{eqnarray}
    &&\rho_i^{(0)}(\mathbf r, \mathbf R_i(t),\theta_i,\beta_2^{(i)})=\rho_0\times \cr &&\left[1+\exp\left(\frac{|\mathbf r-\mathbf R_i(t)|-R_{0i}[1+\beta_2^{(i)}Y_{20}(\theta_i)]}{a_0}\right)\right]^{-1}
\end{eqnarray}
where $\rho_0$=0.17 fm$^{-3}$ and $a_0=0.54$~fm, $\mathbf R_i$ are the center of mass coordinates and $R_{0i}$ are the half density radii of the interaction nuclei.

\bibliography{references}

\end{document}